\documentclass{aa}  

\usepackage{makecell}
\usepackage{graphicx}
\usepackage{natbib}
\usepackage[utf8]{inputenc}
\usepackage{natbib}
\usepackage[mathscr]{euscript}
\usepackage{lscape}
\begin{document}

  \title{Zinc abundances in the Sculptor dwarf spheroidal galaxy\thanks{Based on observations made with ESO/VLT/FLAMES at the La Silla Paranal observatory under program ID 092.B-0194(A)}$^,$}

   \author{\'{A}. Sk\'{u}lad\'{o}ttir
                \inst{1,2}
        \and            
        E. Tolstoy\inst{2}
        \and
          S. Salvadori\inst{3,4,5}    
         \and     
        V. Hill \inst{6}
        \and
         M. Pettini \inst{7}  
          }
          
   \institute{
   			Max-Planck-Institut f$\ddot{\text{u}}$r Astronomie, K$\ddot{\text{o}}$nigstuhl 17, D-69117 Heidelberg, Germany\\
   			\email{skuladottir@mpia.de}
   			\and
              Kapteyn Astronomical Institute, University of Groningen, PO Box 800, 9700AV Groningen, the Netherlands
                \and
                Dipartimento di Fisica e Astronomia, Universita’ di Firenze, Via G. Sansone 1, Sesto Fiorentino, Italy
                \and 
                INAF/Osservatorio Astrofisico di Arcetri, Largo E. Fermi 5, Firenze, Italy
                \and
                GEPI, Observatoire de Paris, PSL Research University, CNRS, Place Jule Janssen 92190, Meudon, France
			\and               
             Laboratoire Lagrange, Universit\'{e} de Nice Sophia Antipolis, CNRS, Observatoire de la C\^{o}te d’Azur, CS34229, 06304 Nice Cedex 
            		4, France
                \and
                Institute of Astronomy, Madingley Road, Cambridge CB3 0HA, England.
                }

\abstract{
From ESO VLT/FLAMES/GIRAFFE spectra, abundance measurements of Zn have been made in $\approx$100 individual red giant branch (RGB) stars in the Sculptor dwarf spheroidal galaxy. This is the largest sample of individual Zn abundance measurements within a stellar system beyond the Milky Way. In the observed metallicity range, $-2.7\leq\text{[Fe/H]}\leq-0.9$, the general trend of Zn abundances in Sculptor is similar to that of $\alpha$-elements. That is, super-solar abundance ratios of [Zn/Fe] at low metallicities, which decrease with increasing [Fe/H], eventually reaching subsolar values. However, at the higher metallicities in Sculptor, $\text{[Fe/H]}\gtrsim-1.8$, we find a significant scatter, $-0.8\lesssim\text{[Zn/Fe]}\lesssim+0.4$, which is not seen in any $\alpha$-element. Our results are consistent with previous observations of a limited number of stars in Sculptor and in other dwarf galaxies. These results suggest that zinc has a complex nucleosynthetic origin, behaving neither completely like an $\alpha$- nor an iron-peak element.
}

   \keywords{Stars: abundances --
                                Galaxies: dwarf galaxies --
                                Galaxies: individual (Sculptor dwarf spheroidal) --
                                Galaxies: abundances --
                                Galaxies: evolution
               }

   \maketitle

%
\section{Introduction}

The chemical evolution pathway of a galaxy is preserved in the photospheres of its long-lived, low-mass stars. Studying chemical abundances of stars from different stages of a galaxy's evolution provides vital information about the processes that dominated in the production of each element. Early in the history of any system, the chemical enrichment of the surrounding interstellar medium (ISM) is dominated by Supernovae (SNe) Type~II, which produce large amounts of $\alpha$-elements (e.g. Mg, Si, S, Ca), and thus high [$\alpha$/Fe]. Typically $\gtrsim$ 1 Gyr after the onset of star formation, SNe Type~Ia start to pollute the environment predominantly with iron-peak elements, and the [$\alpha$/Fe] ratios in the ISM begin to decrease at a rate that depends on the star formation in the system and how it varies with time (e.g. \citealt{MatteucciBrocato1990,GilmoreWyse1991}). 

The element Zn is the heaviest of the iron group. Its nucleosynthetic origin, however, appears to be quite complex and a complete picture for the production has not yet been established. Studies of the solar neighbourhood have demonstrated a flat, moderately enhanced, [Zn/Fe] ratio over a broad metallicity range, $-2.5 \lesssim \textrm{[Fe/H]} \lesssim 0$ (e.g. \citealt{Reddy2003,Reddy2006,Bensby2014}). These results have prompted the use of Zn as a proxy for Fe in Damped Lyman-$\alpha$ systems (DLAs) (e.g. \citealt{Pettini1990}). This is because, unlike Fe, Zn is generally not significantly depleted onto dust in the ISM (e.g. \citealt{SpitzerJenkins1975,Savage1996,Jenkins2009,Vladilo2011}). However, recent observations have shown that Zn can also show a behaviour that is more akin to that of $\alpha$-elements. 

At low metallicities, $\text{[Fe/H]}\leq -2.5$, stars in the Milky Way halo exhibit an enhanced, and possibly metallicity dependent [Zn/Fe] ratio \citep{Primas2000,Cayrel2004}. Furthermore, in the metallicity range $-1.6 \lesssim \textrm{[Fe/H]} \lesssim -0.6$, \citet{NissenSchuster2010,NissenSchuster2011} observed two distinct halo populations in the solar neighbourhood, showing high and low [$\alpha$/Fe] abundance ratios. In their sample, [Zn/Fe] also showed high and low values, similar to the $\alpha$-element variations. Recent results from the Gaia ESO Survey also show some discrepancies from the classical view that Zn follows Fe \citep{Duffau2017}. At lower metallicities, $\text{[Fe/H]}<-0.5$, and at relatively large Galactocentric distances, $R_\textsl{GC}>7$, the observations are consistent with previous measurements in the Milky Way disk, namely that the mildly enhanced values of [Zn/Fe] decrease towards solar values at $\text{[Fe/H]}=0$. At smaller Galactocentric distances, however, a decreasing trend of [Zn/Fe] is observed, reaching sub-solar values, and including some spread. These results are, however, limited to observations of giant stars, since the sample did not include dwarf stars at smaller Galactocentric distances. This matches results for high-metallicity RGB stars at $\text{[Fe/H]}\geq-0.1$ in the Milky Way bulge, where [$\alpha$/Fe] ratios are typically lower compared to the disk, and show a spread of $-0.60<\text{[Zn/Fe]}<+0.15$, with most stars having subsolar abundance ratios \citep{Barbuy2015}. However, these low values are not observed in the microlensed dwarf and subgiant stars in the Galactic bulge \citep{Bensby2013}, and the discrepancy between these two surveys has not yet been resolved. Analogous to what is observed in the Milky Way, previously published [Zn/Fe] abundance ratio measurements in Sculptor and other dwarf galaxies show low values, and some spread (e.g. \citealt{Shetrone2001,Shetrone2003,Sbordone2007,Cohen2010,Venn2012,Berg2015}).

Overall, the observational findings are in agreement with theoretical calculations, which predict the [Zn/Fe] ratios in the yields of Type Ia SNe to be negative (e.g. \citealt{Iwamoto1999,Kobayashi2006}), and lower than in Type II SNe, similar to $\alpha$-elements. However, the observed trend of [Zn/Fe] with [Fe/H] in the Milky Way cannot be fully explained only with a mixture of normal core-collapse SNe Type II and Type Ia. To reproduce the high values observed at the lowest metallicities ($\text{[Zn/Fe]}\sim+0.5$, at $\text{[Fe/H]}<-3$), \citet{UmedaNomoto2002} suggest that SNe with higher explosion energies, so called hypernovae (HNe), are required. Similarly, \citet{Kobayashi2006} suggest that a significant contribution of HNe is needed to reproduce the values observed in the solar neighbourhood, namely $\text{[Zn/Fe]}=0$ at $\text{[Fe/H]}=0$. In addition to classical SNe explosions, other production sites have been proposed. \citet{Hoffman1996} predicted a significant contribution of Zn production from $\alpha$-rich neutrino-driven winds, following the delay of supernovae explosions. The weak $s$-process in massive stars and the main $s$-process occurring in AGB stars are also predicted to account for a total of $\approx$11\% of the Zn in the solar neighboorhood \citep{Travaglio2004}.


\begin{table}
\caption{Log of the VLT/FLAMES service mode observations.}
\label{tab4:obs}
\centering
\small
\begin{tabular}{c c c c c}
\hline\hline
Date    &       Plate   &       Exp.time        &       Airmass &       Seeing  \\
        &               &        (min)  &       (average)       &       (arcsec)        \\
\hline
2013-Oct-08	&	MED2      	&	60.00	&	1.09	&	1.5       	\\
2013-Oct-30	&	MED1      	&	56.25	&	1.23	&	1.6       	\\
2013-Nov-01	&	MED2      	&	56.25	&	1.09	&	1.3       	\\
2013-Nov-27	&	MED1      	&	56.25	&	1.07	&	1.4       	\\
2013-Nov-27	&	MED1      	&	56.25	&	1.17	&	1.4       	\\
2013-Nov-28	&	MED2      	&	56.25	&	1.10	&	1.5       	\\
\hline
\end{tabular}
\end{table}

On the other hand, very massive stars, $140 < M_\star/M_\odot < 260$, which end their lives as Pair Instability Supernovae (PISN), are expected to produce significant amounts of Fe and the $\alpha$-elements but only a negligible amount of Zn, with yields as low as $\text{[Zn/Fe]} < -1.5$ (e.g. \citealt{UmedaNomoto2002,HegerWoosley2002,Kozyreva2014}). Such a stellar population might have been  abundant in the early Universe because the initial mass function (IMF) of the first stars was likely top heavy (e.g. \citealt{Hosokawa2011,Hirano2014}). The chemical signature of these zero-metallicity PISN might be preserved in long-lived, relatively metal-rich stars at $\text{[Fe/H]} \gtrsim -2$ \citep{Salvadori2007,deBennassuti2017}, which should show low [Zn/Fe] values accompanied with a high abundance ratios between even and odd elements (e.g. \citealt{HegerWoosley2002}).

To get a global overview of the apparently complicated production of Zn, it is important to provide observational constraints in as many different environments as possible. The Sculptor dwarf spheroidal galaxy (dSph) in the Local Group is one of the few extragalactic systems where we are able to obtain an unobscured picture of the early star formation and chemical enrichment. This ancient galaxy had a simple star formation history, with a peak in star formation $\sim$13~Gyr ago and a slow decrease, so the majority of the stars were formed during the first 2-3~Gyr \citep{deBoer2012}. This galaxy is dominated by an old stellar population (>10~Gyr), and thus it gives us a clear view back to the star formation and chemical enrichment processes in the early Universe. 

In this paper, we present Zn abundance determinations for $\approx$100 red giant branch (RGB) stars in Sculptor, expanding significantly the data base of such measurements in Local Group galaxies. Taking into account the old age of the stellar population in Sculptor, these results are directly complementary to chemical abundance studies of DLA systems at high redshift. A detailed comparison between chemical abundances in local dwarf galaxies and DLA systems will be the subject of an upcoming paper, Skúladóttir et al. in prep.

%
\section{Observations and data reduction}

The observations were taken in service mode with VLT FLAMES/GIRAFFE in October and November of 2013, using the HR7A grating, which covers the wavelength range 4700-4970 \AA\ with resolution R$\sim$19,500. The observational details are listed in Table \ref{tab4:obs}. 

In the $25^\prime$ diameter field placed on the centre of the Sculptor dSph, detailed abundance measurements have been made for $\approx$100 stars (\citealt{Tolstoy2009}, Hill et al in prep.). For an overlapping sample of 86 stars, high-resolution (HR) FLAMES/GIRAFFE spectroscopy has previously been carried out to measure S \citep{Skuladottir2015b}, and now the same stars were targeted with the HR7A grating to measure Zn. At the distance of Sculptor, only the brightest stars are available for HR spectroscopy. This sample, therefore, consists of RGB stars, with $T_\textsl{eff} \lesssim 4700$~K. 

   \begin{figure}
   \centering
   \includegraphics[width=\hsize-1.cm]{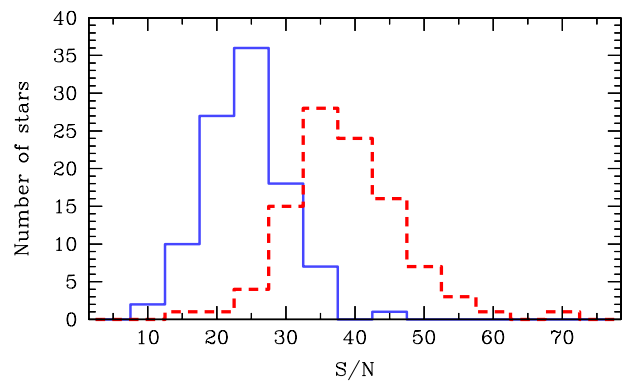}
      \caption{Signal-to-noise, per pixel, achieved for all the stars in the sample, in the regions 4750-4850~\AA\ (blue solid line), and 4850-4950~\AA\ (red dashed line).
      }
         \label{fig:sn}
   \end{figure}

The GIRAFFE spectra were reduced with the ESO-provided pipeline\footnote{ftp://ftp.eso.org/pub/dfs/pipelines/giraffe/giraf-pipeline-manual-2.14.pdf}, including bias, flat-field, and wavelength calibration and extraction. Each observation was reduced separately via the SUM method, provided by the pipeline. The final reduced sets of spectra were sky-subtracted using a routine written by M. Irwin (see \citealt{Battaglia2008b}), which scales the sky background to be subtracted from each object spectrum to match the observed sky emission lines. Each set of spectra was combined using a weighted mean of the counts going into each observation, excluding pixels with extreme outliers in individual exposures. 

The signal-to-noise ratio, S/N, was evaluated as the mean value over the standard deviation of the continuum in line-free regions. The flux is low in our sampled RGB stars, especially in the blue. Due to the relatively steep slope of the flux (because of changes both in the luminosity of the star and instrument efficiency) the S/N was measured in two parts of the spectra, in the region 4750-4850~\AA, where we measured Zn, and in the red part, 4850-4950~\AA. This is shown in Fig.~\ref{fig:sn} and listed in Table~2. The S/N at the bluest end of the spectra, 4700-4750~\AA\ was very low ($\lesssim$10), and in general this region is not usable for accurate abundance measurements, with the exception of the brightest stars.

      \begin{figure}
   \centering
   \includegraphics[width=\hsize-1.cm]{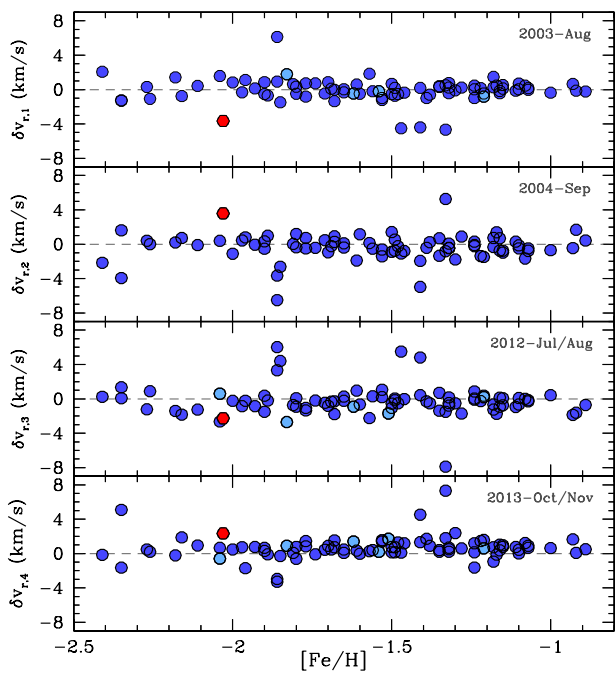}
      \caption{Deviation from the average value of the stellar velocities, $\delta v_{r,i}=v_{r,i}-\bar{v}_{r}$, for a series of measurements. Red hexagon is ET0097, the only known CEMP-no star in Sculptor \citep{Skuladottir2015a}, and light blue circles are stars with less than four velocity measurements. From top to bottom: Hill et al. (in prep.); \citet{Battaglia2008b}; \citet{Skuladottir2015b}; this work.
      }
         \label{fig:vrcomp}
   \end{figure}

%
\section{Velocity measurements}

A large fraction of our target stars have previously been observed in other wavelength regions: with the low-resolution setting LR8, $R\sim 6,500$, in September 2004 \citep{Battaglia2008b}; high-resolution settings HR10, $R\sim20,000$, in August 2003 (Hill et al. in prep.); and HR22B, $R\sim20,000$, in July and August 2012 \citep{Skuladottir2015b}. These observations span nine years, and the velocity measurements of individual stars, see Table~\ref{tab4:vr}, can be used to identify those with velocity variations, i.e. stars that are likely to have a binary companion. In all cases the velocities are measured using a cross-correlation with an RGB template. The results from \citet{Battaglia2008b} are obtained with a single exposure, while the HR results are taken from multiple exposures either over few days (Hill et al. in prep.), or approximately a month (\citealt{Skuladottir2015b}; this work), and the average of the exposures is listed. On these short timescales, $\lesssim$ 1 month, none of the stars within the samples showed a significant velocity variation, beyond what can be expected from measurement errors.

How the velocity measurements, taken over years, differ from the mean for each star, $\delta v_{r,i}=v_{r,i}-\bar{v}_{r}$, is shown in Fig.~\ref{fig:vrcomp}. The total error of the velocity values is the combination of random measurement errors due to noise, biases because of the difference in templates used for the cross-correlation, and uncertainties in the wavelength calibration of each observed FLAMES setting. This is best calculated over the whole sample. The median offset in each case is:  $\delta\widetilde{v}_{r,1}=0.0$ (Hill et al. in prep.), $\delta\widetilde{v}_{r,2}=-0.2$~km/s \citep{Battaglia2008b}, $\delta\widetilde{v}_{r,3}=-0.3$~km/s \citep{Skuladottir2015b}, and $\delta\widetilde{v}_{r,4}=+0.6$~km/s (this work). We note that the intrinsic error of the velocity measurements are highest for the LR setting, $v_{r,2}$, on the order of few km/s, while for the HR settings the measurement error is $\lesssim1$~km/s.

   \begin{figure}
   \centering
   \includegraphics[width=\hsize-1.cm]{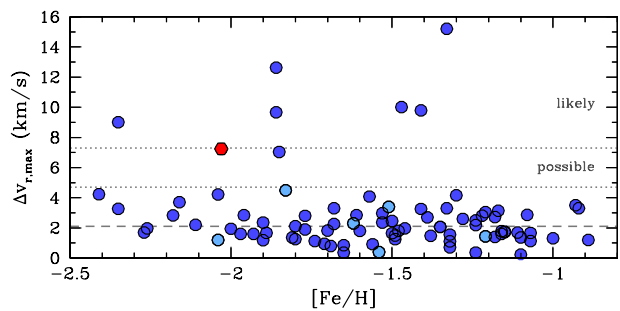}
      \caption{Maximum velocity variation for each star, between the four different measurements. Dashed line shows the median over the sample, and dotted lines show $1\sigma$ and $2\sigma$ variations from this value, marking where possible and likely binaries lie, respectively. Symbols are the same as in Fig.~\ref{fig:vrcomp}.
      }
         \label{fig:vr}
   \end{figure}

The maximum velocity variations for each star between these four measurements, $\Delta v_{r,max}$, are shown in Fig.~\ref{fig:vr}. The median value over the whole sample is $\Delta \bar{v}_{r,max}=2.1$~km/s, which arises mostly due to errors, with a standard deviation of $\sigma=2.6$~km/s. The stars that show a variation that is 1$\sigma$ and $2\sigma$ greater than the median value, $\Delta \widetilde{v}_{r,max}$, are labelled possible and likely binaries respectively in Table~\ref{tab4:vr}. Out of these, all would still be considered possible/likely binaries if measurements from the LR setting, $v_{r,2}$ were excluded.

In our sample, 8 stars out of 87 are possible or likely binaries, that is 9\%. This should be considered a lower limit on the binary fraction in this sample, as the number of velocity measurements is limited. In particular, we note that ET0097, the only carbon-enhanced metal-poor (CEMP) star in the sample \citep{Skuladottir2015a} is a possible binary. This star shows no Ba enhancement, and is therefore a CEMP-no star, which are not typically binaries, though some of them are \citep{Starkenburg2014,HansenT2015b}. In the case of ET0097, its carbon enhancement cannot be easily explained with mass transfer from a binary companion as there are no clear signs of this in the abundance patterns \citep{Skuladottir2015a}.

\setcounter{table}{4}
  \begin{table*}
\caption{Photometry of the new targets and ET0147.}
\label{tab4:phot}
\centering
\small
\begin{tabular}{ccccccccccccc}
\hline\hline
Star	&	B	&	$\delta_\text{B}$	&	V	&	$\delta_\text{V}$	&	I	&	$\delta_\text{I}$	&	J	&	$\delta_\text{J}$	&	H	&	$\delta_\text{H}$	&	K	&	$\delta_\text{K}$	\\
\hline
ET0034	&	18.545	&	0.007	&	17.490	&	0.003	&	16.309	&	0.007	&	15.509	&	0.004	&	14.965	&	0.004	&	14.860	&	0.006	\\
ET0135	&	18.380	&	0.003	&	17.070	&	0.003	&	15.648	&	0.006	&	14.678	&	0.002	&	14.053	&	0.002	&	13.922	&	0.003	\\
ET0156	&	18.911	&	0.003	&	17.620	&	0.003	&	16.277	&	0.007	&	15.318	&	0.003	&	14.688	&	0.003	&	14.536	&	0.005	\\
ET0223	&	19.353	&	0.003	&	18.350	&	0.002	&	17.218	&	0.005	&	16.384	&	0.007	&	15.873	&	0.007	&	15.764	&	0.011	\\
ET0332	&	19.074	&	0.003	&	18.070	&	0.002	&	16.944	&	0.007	&	16.116	&	0.006	&	15.602	&	0.006	&	15.490	&	0.009	\\
ET0351	&	19.477	&	0.004	&	18.300	&	0.002	&	17.102	&	0.006	&	16.182	&	0.006	&	15.591	&	0.006	&	15.459	&	0.009	\\
ET0359	&	19.351	&	0.005	&	18.410	&	0.004	&	17.326	&	0.006	&	16.540	&	0.008	&	16.058	&	0.008	&	15.937	&	0.012	\\
ET0375	&	18.863	&	0.004	&	17.970	&	0.004	&	16.918	&	0.006	&	16.172	&	0.006	&	15.695	&	0.006	&	15.587	&	0.009	\\
ET0388	&	19.601	&	0.014	&	18.140	&	0.006	&	16.959	&	0.007	&	16.106	&	0.006	&	15.578	&	0.006	&	15.483	&	0.009	\\
ET0393	&	19.259	&	0.003	&	18.170	&	0.003	&	17.092	&	0.007	&	16.236	&	0.006	&	15.704	&	0.006	&	15.591	&	0.009	\\
ET0394	&	19.309	&	0.003	&	18.160	&	0.003	&	16.963	&	0.006	&	16.042	&	0.006	&	15.435	&	0.006	&	15.313	&	0.008	\\
ET0396	&	19.240	&	0.004	&	18.210	&	0.002	&	17.095	&	0.006	&	16.256	&	0.006	&	15.715	&	0.006	&	15.633	&	0.010	\\
ET0398	&	19.376	&	0.003	&	18.200	&	0.003	&	16.992	&	0.007	&	16.052	&	0.006	&	15.450	&	0.006	&	15.321	&	0.008	\\
ET0402	&	19.378	&	0.003	&	18.270	&	0.002	&	17.096	&	0.006	&	16.211	&	0.006	&	15.640	&	0.006	&	15.521	&	0.009	\\
ET0408	&	19.437	&	0.004	&	18.380	&	0.003	&	17.244	&	0.005	&	16.415	&	0.007	&	15.858	&	0.007	&	15.747	&	0.011	\\
ET0147	&	18.795	&	0.003	&	17.459	&	0.002	&	16.083	&	0.005	&	15.078	&	0.003	&	14.431	&	0.003	&	14.268	&	0.004	\\
\hline
\end{tabular}
\end{table*}

   %
 \section{Abundance analysis}

Our analysis was carried out using the spectral synthesis code TURBOSPEC\footnote{ascl.net/1205.004} \citep{AlvarezPlez1998,Plez2012}. The stellar atmosphere models are adopted from MARCS\footnote{marcs.astro.uu.se} \citep{Gustafsson2008} for stars with standard composition, 1D and assuming LTE, interpolated to match the exact stellar parameters for the target stars. Atomic parameters are adopted from the VALD\footnote{http://vald.astro.uu.se} database (\citealt{Kupka1999} and references therein). All lines used for abundance measurements are listed in Table~\ref{tab4:linelisti}. All measurements are done by including all atomic data for the wavelength range in question, thus including blends of other elements. A single broadening factor was added, corresponding to the width of which is the quadratic sum spectrograph resolution and macro turbulence, calibrated against the available, unblended Fe I lines in the spectra. To be consistent with previous work on the same stellar sample, the adopted solar abundances are from \citet{GrevesseSauval1998}; A(Fe)$_\odot=7.50$, A(Zn)$_\odot=4.60$, and A(Ti)$_\odot=5.02$. Literature data used in this paper are scaled to match these solar abundances.

\subsection{Stellar parameters}

The stellar parameters ($T_\textsl{eff}$, $\log g$, and $v_t$) and [Fe/H] for most of the target stars were previously determined by Hill et al. (in prep.), and are the same as used in \citet{Skuladottir2015b}. The data used for this include FLAMES/GIRAFFE settings HR10, HR13, HR14A and HR15 and two stars overlapping with our sample have FLAMES/UVES spectra. The stellar parameters for these stars were determined by following a method described in \citet{Letarte2010} for the GIRAFFE sample, and in \citet{Shetrone2003} for the UVES stars.  The spectra used in Hill et al. (in prep.) cover longer wavelength and suffer less crowding and blending of lines, compared to the HR7A used here. We therefore choose to use these previously determined stellar parameters where possible, both for higher quality and for consistency with previous abundance measurements of this sample (Hill et al. in prep; \citealt{Tolstoy2009,Skuladottir2015b}). For new targets, however, we determine the stellar parameters with the available HR7A spectra. All adopted stellar parameters are listed in Table~2.

To make sure that the abundance determinations are all on the same scale, [Fe/H] was compared in the overlapping 86 stars. To check that the available Fe lines were reliable, we first performed a selection based on the statistical behaviour of the lines. Out of the $\approx50$ lines in the wavelength range, in total 39 Fe lines were selected and measured in stars for this comparison. This was done where the S/N of the spectrum was sufficient for the given line, and it was not too severely blended. The total number of Fe lines used for each star ranged from 13 to 37, depending on the metallicity of the star, and the quality of the spectra. 

\begin{figure}
   \centering
   \includegraphics[width=\hsize-1.cm]{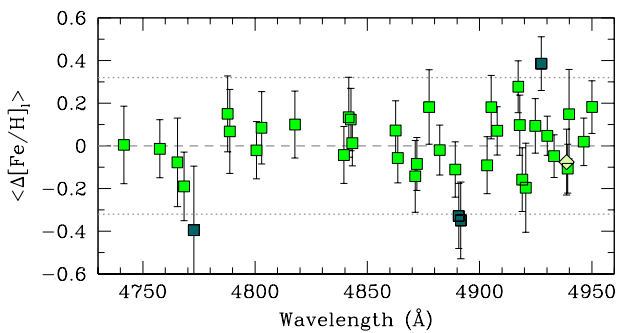}
      \caption{Average deviation of the measured Fe lines from the mean, as a function of wavelength. The standard deviation of the sample of stars is shown by an errorbar. The dotted lines show the 2$\sigma$ interval of the scatter. Green squares are Fe~I lines, while the one available Fe~II line is a pale green diamond. Dark green squares are Fe~I lines that deviate more than 2$\sigma$ from the mean. We note that a different number of measurements goes into each point, depending on the number of stars in which the line could be measured. 
      }
         \label{fig:Felines}
   \end{figure}

\begin{figure}
   \centering
   \includegraphics[width=\hsize-1.cm]{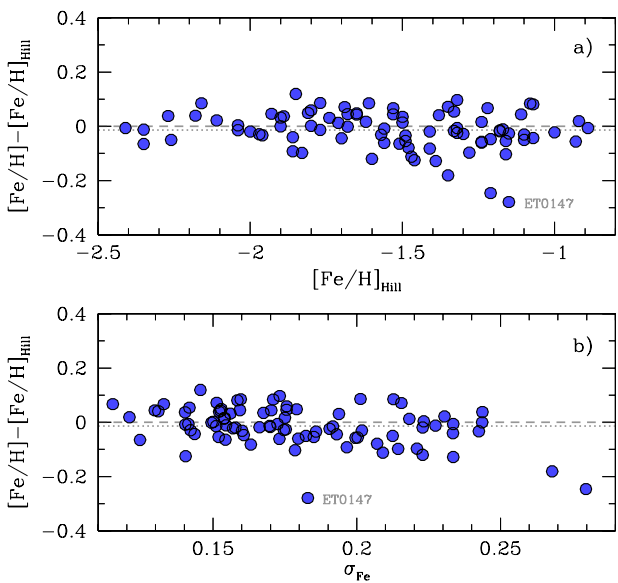}
      \caption{Difference between [Fe/H] measurements from this work, and Hill et al. (in prep.) as a function of: a) [Fe/H]$_\textsl{Hill}$; b) the scatter among Fe lines, $\sigma_\text{Fe}$, as measured here. The dotted line shows the average difference between the two measurements.
      }
         \label{fig:sFediff}
   \end{figure}

For every star and for each line $l$ we measured the deviation of the [Fe/H]$_l$ measurement from the average, $\Delta \text{[Fe/H]}_l=\text{[Fe/H]}_{l}-\text{$<$[Fe/H]>}$. The mean value of this difference, over the sample of stars for each line, $\text{<}\Delta\text{[Fe/H]}_l\text{>}$, is shown in Fig.~\ref{fig:Felines}. Lines that deviate more than 2$\sigma$ from the average, are deemed unreliable and are excluded from the final analysis. The reason for the large deviation in these lines is most likely either blending that is not correctly accounted for in the synthetic spectra, or incorrect atomic parameters. The comparison of [Fe/H] is shown in Fig.~\ref{fig:sFediff}. Our [Fe/H] measurements are systematically only $0.01\pm0.01$~dex lower than the results obtained from Hill et al. (in prep), so we conclude that the two measurements are on the same abundance scale.

Two stars in the sample with $\text{[Fe/H]}_\textsl{Hill}\approx-1.2$, have $\text{[Fe I/H]}_\textsl{HR7}-\text{[Fe/H]}_\textsl{Hill}<-0.2$~dex. One of these stars, ET0342, has the largest standard deviation between individual Fe lines in the sample, $\sigma_\text{Fe}=0.28$, which means that the measured Fe lines do not agree well with each other. This spectrum also has the lowest S/N of all target stars, and so the quality of the spectrum is not good enough for reliable abundance measurements and we exclude it from further analysis. The other star, ET0147, has  $\sigma_\text{Fe}=0.18$, which is quite typical for the target stars, yet the [Fe/H] measurements show a large difference. This star, has a very low S abundance and, as noted in \citet{Skuladottir2015b}, it has the biggest difference between $T_\textsl{eff}$, as determined from photometry and spectroscopy, in the entire sample of Hill et al. (in prep.). Therefore, we will re-evaluate the stellar parameters of ET0147 along with the new targets in the next section.

\subsubsection{New targets}

In addition to the targets overlapping with Hill et al. (in prep.), 15~new RGB stars were observed. The stellar parameters for these stars, and the star ET0147, were determined from photometry (using B, V, I, J, H and K filters), and the HR7A spectra. The photometry for these 16 stars is listed in Table~\ref{tab4:phot} with errors, and comes from \citet{deBoer2011} and VISTA archival observations (M. Irwin, private communication).

The effective temperature, $T_\textsl{eff}$, was determined with photometry, following the recipe from \citet{RamirezMelendez2005}, assuming reddening correction, $E(C)$, in the direction of the Sculptor dSph as listed in Table~\ref{tab4:E}. The estimates from different colors, $T_\textsl{eff}$(C), are listed in Table~7. In a few cases the photometry falls out of the range provided by \citet{RamirezMelendez2005}, and $T_\textsl{eff}$(C) is not calculated. The final value $T_\textsl{eff}$ is determined by the average of the available $T_\textsl{eff}$(C). The error is determined by the quadratic sum of relevant factors:

\begin{equation}
\delta_T=\sqrt{ \sigma_T^2 + \delta T_\textsl{eff}(\delta_\text{[Fe/H]})^2}
\end{equation}

\noindent where $\sigma_T$ is the scatter between the available $T_\textsl{eff}$(C), and $\delta T_\textsl{eff}(\delta_\text{[Fe/H]})$ is the error in $T_\textsl{eff}$ arising from the uncertainty in [Fe/H]. In general, $\sigma_T$ is the bigger source of error.

\setcounter{table}{5}
\begin{table}
\caption{Reddening corrections towards the Sculptor dSph from \citet{Schlegel1998}.}
\label{tab4:E}
\centering
\begin{tabular}{c c}
\hline\hline
Colour    &       Red. Corr., E(C)    \\
\hline
B-V	&   0.016	  \\
V-I	&	  0.023   \\
V-J	&	  0.037   \\
V-H	&	  0.042   \\
V-K	&	  0.045   \\
\hline
\end{tabular}
\end{table}

The surface gravities are obtained using photometry and the standard relation:
\begin{equation}
\log g_{\star}=\log g_{\odot}+\log{\frac{\text{M}_{\star}}{\text{M}_{\odot}}}+  4\log{ \frac{T_{\textsl{eff,}\star}}{T_{\textsl{eff,}\odot}} }+0.4(M_{\textsl{bol,}\star}-M_{\textsl{bol,}\odot})
\end{equation}
The absolute bolometric magnitudes for the stars, $M_{\textsl{bol,}\star}$ are calculated using a calibration for the V-band magnitude \citep{Alonso1999} and a distance modulus of $(m-M)_0~=~19.68\pm0.08$, from \citet{Pietrzynski2008}. The mass of each star is assumed to be $\text{M}_\star~=~0.8\pm0.2 $~M$_\odot$, in agreement with the star formation history of Sculptor \citep{deBoer2012}. The solar values used are: $\log~g_\odot=4.44$, $T_{\textsl{eff,}\odot}~=~5790$~K and $M_{\textsl{bol,}\odot}~=~4.72$ (to keep consistency with \citealt{Starkenburg2013,Skuladottir2015a}). The errors in the surface gravities are determined as the quadratic sum of the errors due to each variable in the equation (where the uncertainty in $(m-M)_0$ is the dominating factor), and are listed along with $\log g_{\star}$  in Table~7.

The method used to determine photometric $T_\textsl{eff}$ and $\log g$ in Hill et al. (in prep.) is the same as described here, but a different distance modulus was adopted $(m-M)_0~=~19.54$, in agreement with \citet{Tolstoy2003}, and only colors (V-I), (V-J) and (V-K) were used. Furthermore, the photometric values of $T_\textsl{eff}$ were checked by examining the dependence of Fe I abundances of individual lines on their excitation potential, $\chi$. This lead to the revision of the values in 12 stars in their sample, typically of $\lesssim100$~K. The final values of $\log g$ were determined spectroscopically by demanding that abundances of Fe I and Fe II agree within the uncertainties.

The turbulence velocity, $v_t$ was determined by minimizing the slope of [Fe/H] with $\log$(EW/$\lambda$), in the same way as done in Hill. et al. (in prep). A typical error of 0.2~km/s was adopted, taken from the errors on the slope. Sometimes the available Fe~I lines were too few to robustly determine this slope. In these cases, the value $v_t=1.70\pm0.25$~km/s was adopted, since this is the median value of the entire sample as determined by Hill et al. (in prep.) and the associated error was assumed to be the standard deviation. 

\setcounter{table}{6}
\begin{sidewaystable*}
\label{tab:stellpar}
\centering
\caption{Stellar parameters for the new targets and ET0147.}
\begin{tabular}{lcccccccccccccccc}
\hline\hline
Star	&	$T_\textsl{eff}$(B-V)	&	$T_\textsl{eff}$(V-I)	&	$T_\textsl{eff}$(V-J)	&	$T_\textsl{eff}$(V-H)	&	$T_\textsl{eff}$(V-K)	&	$\sigma_T$	&	$T_\textsl{eff}$	&	$\delta_T$	&	$\log g $	&	$\delta_\text{log(g)}$	&	$v_t$	&	$\delta_{v_t}$	&	$N_\text{Fe}$	&	$\sigma_\text{Fe}$	&	[Fe/H]	&	$\delta_\text{[Fe/H]}$	\\
	&	(K)	&	(K)	&	(K)	&	(K)	&	(K)	&	(K)	&		(K)		&	(K)	&		&		&	km/s	&	km/s	&		&		&		\\
\hline		
ET0034	&	4418	&	4495	&	4526	&	4516	&	4540	&	48	&	4499	&	49	&	0.94	&	0.12	&	1.70	&	0.25	&	17	&	0.16	&	-2.41	&	0.09	\\
ET0135	&	4176	&	...	&	4182	&	4163	&	4185	&	10	&	4176	&	18	&	0.56	&	0.12	&	1.65	&	0.20	&	32	&	0.20	&	-2.32	&	0.06	\\
ET0147$^a$	&	4089	&	4189	&	4075	&	4108	&	4095	&	45	&	4111	&	51	&	0.66	&	0.12	&	1.30	&	0.20	&	30	&	0.16	&	-1.39	&	0.08	\\
ET0156	&	4188	&	4240	&	4186	&	4211	&	4208	&	22	&	4207	&	32	&	0.77	&	0.12	&	1.25	&	0.20	&	42	&	0.17	&	-1.37	&	0.06	\\
ET0223	&	4511	&	4532	&	4499	&	4548	&	4558	&	25	&	4529	&	25	&	1.31	&	0.12	&	1.60	&	0.20	&	26	&	0.23	&	-1.91	&	0.07	\\
ET0332	&	4491	&	4557	&	4530	&	4559	&	4571	&	32	&	4542	&	32	&	1.20	&	0.12	&	1.70	&	0.25	&	21	&	0.26	&	-2.17	&	0.08	\\
ET0351	&	4305	&	4404	&	4286	&	4328	&	4317	&	45	&	4328	&	48	&	1.16	&	0.12	&	1.45	&	0.20	&	29	&	0.19	&	-1.27	&	0.09	\\
ET0359	&	4564	&	4629	&	4627	&	4655	&	...	&	38	&	4619	&	41	&	1.37	&	0.12	&	1.70	&	0.25	&	11	&	0.22	&	-2.67	&	0.11	\\
ET0375	&	4643	&	4691	&	4723	&	4740	&	4751	&	44	&	4710	&	46	&	1.14	&	0.13	&	1.70	&	0.25	&	15	&	0.24	&	-2.28	&	0.10	\\
ET0388	&	...	&	4447	&	4418	&	4476	&	4498	&	35	&	4460	&	41	&	1.19	&	0.12	&	1.70	&	0.25	&	25	&	0.25	&	-1.97	&	0.08	\\
ET0393	&	4382	&	4634	&	4551	&	4561	&	4572	&	37	&	4579	&	39	&	1.26	&	0.12	&	1.35	&	0.20	&	22	&	0.28	&	-2.12	&	0.09	\\
ET0394	&	4336	&	4409	&	4289	&	4316	&	4313	&	46	&	4333	&	48	&	1.11	&	0.12	&	1.50	&	0.20	&	40	&	0.21	&	-1.39	&	0.08	\\
ET0396	&	4489	&	4551	&	4502	&	4531	&	4561	&	31	&	4527	&	33	&	1.24	&	0.12	&	1.55	&	0.20	&	35	&	0.17	&	-1.59	&	0.06	\\
ET0398	&	4310	&	4388	&	4257	&	4297	&	4290	&	49	&	4309	&	49	&	1.11	&	0.12	&	1.55	&	0.20	&	38	&	0.24	&	-1.29	&	0.08	\\
ET0402	&	4397	&	4442	&	4345	&	4389	&	4384	&	35	&	4391	&	38	&	1.19	&	0.12	&	1.40	&	0.20	&	40	&	0.22	&	-1.30	&	0.08	\\
ET0408	&	4452	&	4517	&	4490	&	4509	&	4517	&	28	&	4497	&	30	&	1.28	&	0.12	&	1.40	&	0.20	&	32	&	0.19	&	-1.49	&	0.07	\\
\hline
\multicolumn{7}{l}{$^a$ From Hill et al. (in prep.), reanalysed.}\\
\end{tabular}
\end{sidewaystable*} 

The [Fe/H] of these stars was determined using the available Fe~I lines in the wavelength region 4740-4970~\AA. In total 54 lines were used that did not systematically show greater than 2$\sigma$ deviation from the mean, see Table~\ref{tab4:linelisti}. Some of these lines were only suitable for metal-poor or metal-rich stars, depending on strength and blending, so the number of lines used for each target, $N_\text{Fe}$, ranges from 11 to 42. The results are shown in Table~7, where the error on [Fe/H] is:

\begin{equation}
\delta_\text{Fe}=\sqrt{ \delta_\textsl{noise}^2  + \Delta \text{[Fe/H]}_\text{sp}^2  }    
\end{equation}
\noindent where $\Delta \text{[Fe/H]}_\text{sp}$ is the uncertainty coming from the stellar parameters and: 
\begin{equation}
\delta_\textsl{noise}=  \frac{\sigma_\text{Fe}}{\sqrt{N_\text{Fe}-1} }
\end{equation}
The metallicities of all the new stars lie within the range $-2.7~\leq~\text{[Fe/H]}~\leq~-1.3$.

       \begin{figure}
   \centering
   \includegraphics[width=\hsize-1.cm]{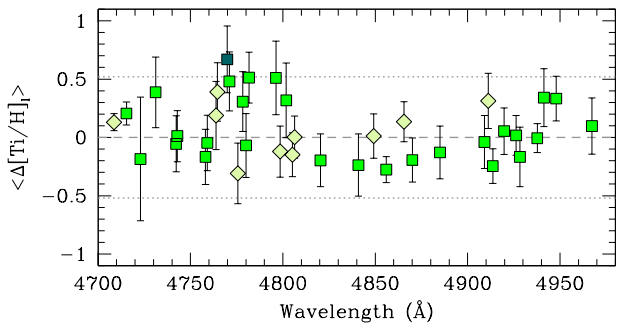}
      \caption{Average deviation of the measured Ti lines from the mean, as a function of wavelength, equivalent to Fig.~\ref{fig:Felines}. Green squares are Ti~I lines, pale green diamonds are Ti~II lines and the dark green square is a Ti~I line falling out of the criteria. The bluest part of the spectrum ($<4740$~\AA) is only usable for the brightest stars, so these points and their standard deviations are determined from measurements in $\leq 6$ stars.        
      }
         \label{fig:Tilines}
   \end{figure}


\subsection{Titanium measurements}
The wavelength region observed with the HR7A setting has not commonly been used for faint targets, such as RGB stars in dwarf galaxies. To ensure that the data reduction was successful, and to have a reference element for scaling the error estimates of Zn (see Section 4.4), Ti abundances were measured for the sample. The spectra contain $\sim$40 measurable Ti lines, given the resolution, the S/N ratios and the stellar parameters of the sample. Ti abundances were measured, using 38 lines, of both Ti~I and Ti~II, which are shown in Fig.~\ref{fig:Tilines} and listed in Table~\ref{tab4:linelisti}. One of these lines was excluded, since it showed a systematic deviation from the average that was more than 2$\sigma$ from the mean. The scatter between lines is larger than for Fe~I lines, because a larger fraction of the lines are at bluer wavelengths ($\leq$4800~\AA) where the S/N is lower. The fact that both Ti~I and Ti~II lines are measured together is also expected to increase the scatter, but ionization equilibrium is fulfilled within the errors of the data.

\begin{figure}
   \centering
   \includegraphics[width=\hsize-1.cm]{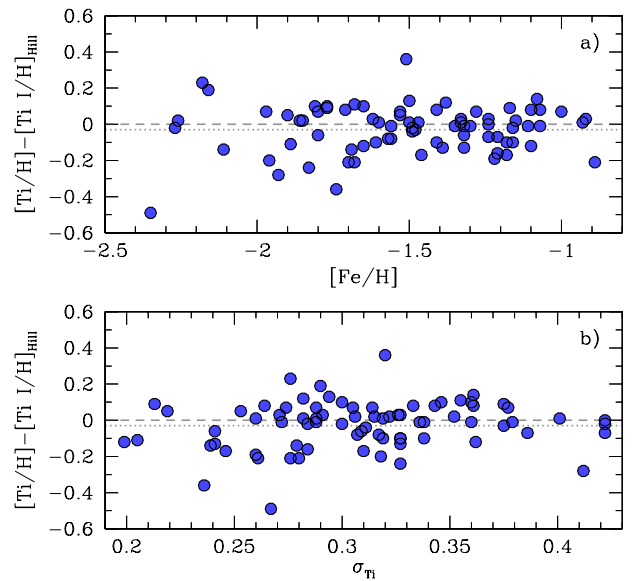}
      \caption{Difference between [Ti/H] from this work, and [Ti~I/H]$_\textsl{Hill}$ from Hill et al. (in prep.) as a function of: a) [Fe/H]; b) the standard deviation between lines, $\sigma_\text{Ti}$, measured here. The dotted lines show the average difference between the two measurements.          
      }
         \label{fig:sTidiff}
   \end{figure}

Hill et al. (in prep.) measured both [Ti~I/H] and [Ti~II/H], and as the measurements here are more dominated by Ti~I lines, a comparison between [Ti/H] and [Ti~I/H]$_\textsl{Hill}$ is shown in Fig.~\ref{fig:sTidiff}. The result obtained here is on average $0.03\pm0.01$ lower than  [Ti~I/H]$_\textsl{Hill}$, and $0.07\pm0.02$ lower than [Ti~II/H]$_\textsl{Hill}$. In most of the sample, the agreement is reasonable, but for a few stars the difference is significant. The measurements from Hill et al. include $\sim$20 Ti I and Ti II lines in total, so in general the measurements made with the HR7A range are expected be more reliable.

\subsection{Zinc measurements}

Two Zn~I lines are available in the observed wavelength range, at 4722.2~\AA\ and 4810.5~\AA\ (see Table~\ref{tab4:linelisti}), but the S/N ratio at the bluest end of the spectra was generally too low for reliable abundance measurements. Even though the bluer Zn~I line could also be measured in the brightest stars, the line at 4810.5~\AA\ was always more reliable, and is therefore used for all stars. Where usable, the line at 4722.2~\AA\ was always in agreement with that at 4810.5~\AA. 

Previous measurements of the Zn abundance in Sculptor have all been made with higher resolution spectra and by using EWs for the abundance determination (\citealt{Shetrone2003,Geisler2005}; \linebreak\citealt{Jablonka2015}; Hill et al. in prep). Here we use synthetic spectra analysis. To ensure the different abundance measurements are in agreement, we used our method on the reduced and normalized spectra from \citet{Shetrone2003} and Hill et al (in prep.) obtained with UVES slit and FLAMES/UVES fibres. Hill et al. (in prep) measured Zn in 7 stars, of which one star is in our sample (ET0112). In total 5 stars in Sculptor were observed by \citet{Shetrone2003}. For one of these stars, ET0158 (H-400), the Zn abundance was not determined, but for the remaining four, three are also in our sample (ET0071, ET0151 and ET0489). 

         \begin{figure}
   \centering
   \includegraphics[width=\hsize-0.5cm]{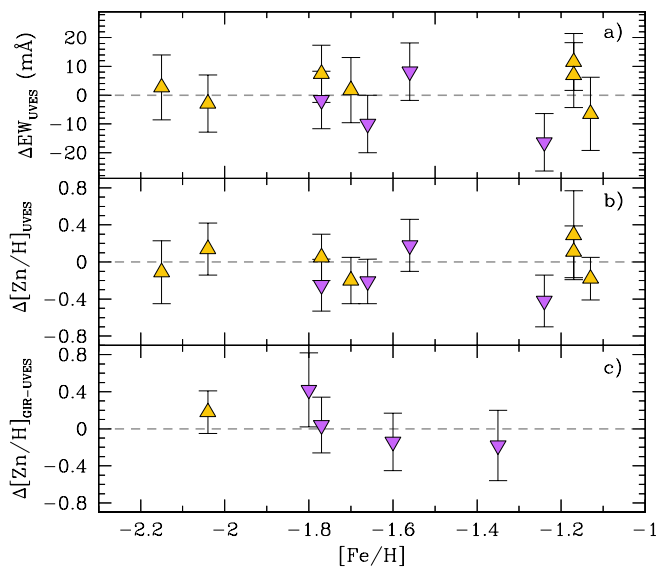}
      \caption{Comparison of our Zn measurements with those from the literature. Errorbars show the combined uncertainties of both measurements. Downward pointing violet triangles are stars from \citet{Shetrone2003} and orange triangles are stars from Hill et al. (in prep.). The panels show differences of:
      a)~The measured EW of the Zn line at 4810.5~\AA, between the literature and this work, in all cases using the same UVES spectra; b)~The [Zn/Fe] measurements, between the literature and this work, in all cases using the same UVES spectra; c)~The [Zn/Fe] measurements from the GIRAFFE HR7A and UVES spectra, as measured by this work.
      }
         \label{fig:EQW}
   \end{figure}

Using the same normalized spectra, the EWs were measured from our best fitting synthetic spectra around the Zn~I line at 4810~\AA, and compared with those from the previous studies see Fig.~\ref{fig:EQW}a. In general the agreement is good, with the exception of ET0071 (H-482) at $\text{[Fe/H]}=-1.35$ ($\text{[Fe/H]}_\text{Shet}=-1.24$), which has a large (negative) spike in the red wing of the line. This may cause errors with both methods but synthetic spectra have the advantage of taking the expected line profile into account. With the same stellar parameters used previously, the Zn abundances were measured using synthetic spectra and compared to the original values, as is shown in Fig.~\ref{fig:EQW}b. Similarly to the previous panel, this comparison is in good agreement with the exception of the star ET0071.

Finally, Fig.~\ref{fig:EQW}c shows the comparison of Zn abundance measurements from GIRAFFE and UVES spectra, as measured by this work. Both continuum evaluation and abundance determination is carried out with synthetic spectra analysis. This panel only contains the stars of the HR7A sample that overlap with Hill et al. (in prep.) and \citet{Shetrone2003}. Contrary to the other panels, ET0158 (H-400), at $\text{[Fe/H]}=-1.80$, is also included  although \citet{Shetrone2003} did not measure the Zn abundance for this star. We use the stellar parameters determined by Hill et al. (in prep.) for the Shetrone stars. All measured stars are in good agreement, whether the UVES or the FLAMES/GIRAFFE spectra are used for the measurements.

\begin{figure}
   \centering
   \includegraphics[width=\hsize-0.5cm]{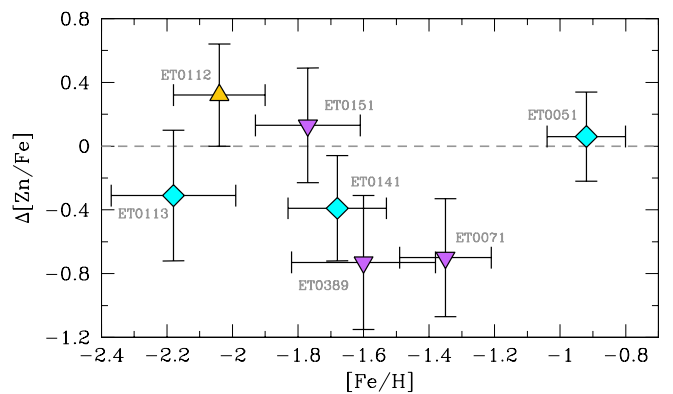}
      \caption{ The difference between [Zn/Fe] measurements from seven stars in this work and the literature. Downward pointing violet triangles are stars from \citet{Shetrone2003}, and cyan diamonds from \citet{Geisler2005}, both using UVES slit spectra. The orange triangle is a star measured by Hill et al. (in prep.) with FLAMES/UVES spectra.
      }
         \label{fig:zncomp}
   \end{figure}

A comparison between our final measured values of [Zn/Fe], and those in the literature for the overlapping sample of stars is shown in Fig.~\ref{fig:zncomp}. We note that two of the stars in common with the \citet{Shetrone2003} sample, ET0071 (H-482) at $\text{[Fe/H]}=-1.35$ and ET0389 (H-459) at  $\text{[Fe/H]}=-1.60$, show a significant difference between the two measurements $\Delta\text{[Zn/Fe]}\approx-0.7$. As previously mentioned, ET0071, has a noisy feature in one of the wings of the Zn~I line, which likely affected the EW in \citet{Shetrone2003}. This is the most likely reason why the EW measurement from the best fit of the synthetic spectra, gives a lower value than measured by \citet{Shetrone2003}, see Fig.~\ref{fig:EQW}a at $\text{[Fe/H]}_\text{Shet}=-1.24$. Regarding the star ET0389, the continuum as evaluated by \citet{Shetrone2003} is higher than that determined here using synthetic spectra. When the same continuum is used, the measurements of [Zn/Fe] agree (see Fig.~\ref{fig:EQW}b, $\text{[Fe/H]}_\text{Shet}=-1.66$). When the method for evaluating continuum that is used here is also used on the UVES spectrum, the two different spectra also agree within the errors (see Fig.~\ref{fig:EQW}c).

Thus, where comparison was possible, the current spectra are in agreement with those of the literature. Different methods of continuum and abundance determination, however, seem to increase the measured scatter. Measuring Zn accurately from only one line at the blue wavelength of 4810.5~\AA\ is challenging in these faint RGB stars, so it is important to keep in mind that the errors in general are quite significant, and should not be neglected.

\subsection{Errors}

In the cases of Fe and Ti, where five or more lines were always measured in the same star, the final abundance was determined to be the average of the measurements, and the error due to the noise was defined as: 

\begin{equation}\label{eq:dnoise}
\delta_{\textsl{noise}} = \frac{\sigma_X}{\sqrt{N_X-1}}
\end{equation}
\noindent where $N_X$ is the number of measured lines of element X, and $\sigma_X$ is the standard deviation of the measurements. 

For Zn only one line was available, and the error for an individual line was determined from the $\chi^2$ fit. The upper and lower errorbars are defined as the values when $\chi^2$ reaches a certain deviation from the best fit
\begin{equation}
\chi^2_\textsl{err}=(1+f)\chi^2_\textsl{bf}
\end{equation}
where $\chi^2_\textsl{bf}$ is the best fit and the constant factor $f=0.35$ is calibrated over the sample so that the average error of Ti is equal to the average dispersion between lines, $\text{<}\delta_\textsl{noise}\text{(Ti)}\text{>}=\text{<}\sigma_\text{Ti}\text{>}$, thereby assuming that the noise is dominating the scatter between Ti lines (both Ti I and Ti II). This $f$ factor is then applied to get the errors for individual Zn lines.
 The error of the line, $\delta_\textsl{noise}$, is taken as the maximum value of the upper and lower errorbars.

The uncertainties of the stellar parameters, $T_\textsl{eff}$, $\log g$, and $v_t$, result in a systematic errors 
in abundance ratios, $\Delta\text{[X/Y]}_\text{sp}$. This is quadratically added to the measurement error, $\delta_{\textsl{noise}}$, to get the final adopted error of abundance ratios:
\begin{equation}\label{eq:totalerr}
\delta_{\text{[X/Y]}}=\sqrt{  \Delta\text{[X/Y]}_\text{sp}^2 +\delta_{\textsl{noise}}(\text{X})^2+\delta_{\textsl{noise}}(\text{Y})^2}
\end{equation}

%
\section{Results}

All abundance measurements are listed with errors in Table~2.

 \subsection{Titanium in Sculptor}

\begin{figure}
   \centering
   \includegraphics[width=\hsize-1.cm]{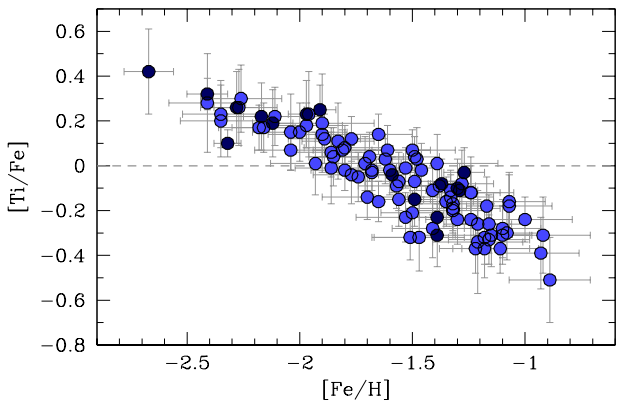}
      \caption{Titanium abundances in Sculptor. Stars overlapping with the sample from Hill et al. (in prep) are blue, while stars observed here for the first time with HR spectra are dark blue.
      }
         \label{fig:TiFe}
   \end{figure}

The trend of decreasing [Ti/Fe] with increasing [Fe/H] in Sculptor is well known (\citealt{Shetrone2003,Geisler2005,Kirby2009,Tolstoy2009}; Hill et al. in prep.). The results of the titanium measurements with the HR7A setting are consistent with this, see Fig.~\ref{fig:TiFe}, and show the typical trend for $\alpha$-elements in dwarf spheroidal galaxies. Less efficient star formation before the onset of SN Type Ia in Sculptor causes the so-called `knee', which is the beginning of the decrease in [$\alpha$/Fe], to occur at a lower [Fe/H] compared to the Milky Way. The same behaviour is also seen in other $\alpha$-elements in Sculptor, for example in sulphur \citep{Skuladottir2015b}.

        \begin{figure}
   \centering
   \includegraphics[width=\hsize-1.cm]{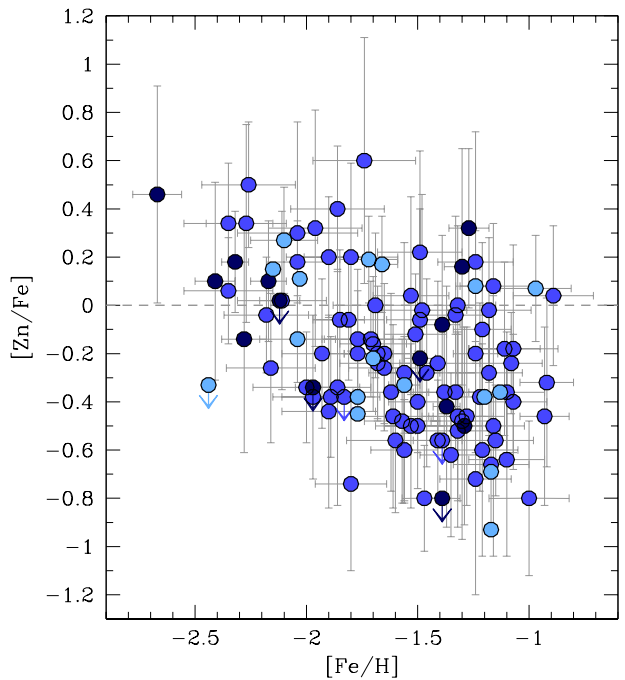}
      \caption{Relation between [Zn/Fe] and [Fe/H]. Symbols are the same as in Fig.~\ref{fig:TiFe} with light blue circles showing results from the literature, using HR UVES data (\citealt{Shetrone2003,Geisler2005,Jablonka2015}; Hill et al. in prep), excluding stars overlapping with the FLAMES sample. Upper limits where $\text{[Zn/Fe]}>0.6$ are not included. 
      }
         \label{fig:ZnFe}
   \end{figure}
   
         \begin{figure}
   \centering
   \includegraphics[width=\hsize-1.cm]{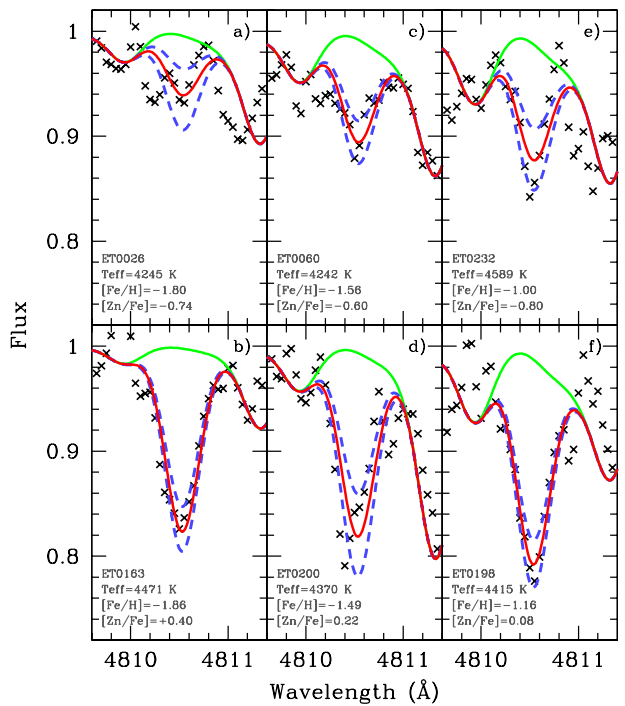}
      \caption{The Zn~I line at 4810.5~\AA\ for three pairs of stars with similar stellar parameters but very different [Zn/Fe] values: a) and b) show stars with $\text{[Fe/H]}\approx-1.8$; c) and d) stars with $\text{[Fe/H]}\approx-1.5$; e) and f) stars with $\text{[Fe/H]}\approx-1.1$. In each case, the spectrum is shown with black crosses, while the best fit is shown with a red solid line, and blue dashed lines show upper and lower errorbars (including error due to stellar parameters). The green solid line shows the case where there is no Zn present.
      }
         \label{fig:Znspec}
   \end{figure}

\subsection{Zinc in Sculptor}
The results of the Zn abundance measurements are shown in Fig.~\ref{fig:ZnFe}. No corrections for non-LTE effects have been applied, but these are expected to be positive and small, $\lesssim~0.1$~dex \citep{Takeda2005}. However, we caution that this conclusion may be biased by the lack of available quantum mechanical data for Zn, for example photo-ionisation cross-sections and inelastic collisions with atomic hydrogen.

The scatter in [Zn/Fe] as a function of [Fe/H] in Sculptor is quite significant. This is also seen in the smaller sample of the literature, obtained from spectra with higher resolution (\citealt{Shetrone2003,Geisler2005,Jablonka2015}; Hill et al. in prep). Over the whole Sculptor sample, the scatter is mostly consistent with the errorbars, which on average are $<\delta_\text{[Zn/Fe]}>=0.34$ (for the FLAMES/GIRAFFE data), while the typical scatter is $\sigma_\text{[Zn/Fe]}\approx0.3$. However, the scatter is not limited to measurements with the largest errorbars, and while all the stars at a given [Fe/H] are consistent with having the same [Ti/Fe] value, the same is not true for [Zn/Fe]. A few examples of stars with similar [Fe/H] and stellar parameters, but clearly different [Zn/Fe] values are shown in Fig.~\ref{fig:Znspec}. Considering the low S/N of our sample, and large errorbars, it is not possible at this point to determine whether there is a real uniform scatter in the data, or if the sample has a narrow trend that includes a few outliers in [Zn/Fe]. 

Low [Zn/Fe] values are of particular interest, since they are one of the signatures of massive pair instability supernovae (PISN). The star ET0026 at $\text{[Fe/H]}=-1.80$, with $\text{[Zn/Fe]}=-0.74\pm 0.32$, see Fig.~\ref{fig:ZnFe} and \ref{fig:Znspec}a, seems to fall out of the typical trend of the other Sculptor stars. This star does not show any peculiar abundances in other $\alpha$- (such as Mg, S, Ca) or iron-peak (such as Ni) elements. Thus, besides low [Zn/Fe], it does not have any signs of the other abundance peculiarities associated with PISN. Another notable low outlier, in Fig.~\ref{fig:ZnFe} is the low upper value of ET0381 from \citet{Jablonka2015} at $\text{[Fe/H]}=-2.4$, with $\text{[Zn/Fe]}<-0.3$. This star has an abnormal abundance pattern, with low values of $\alpha$-elements and other iron-peak values compared to Fe. The peculiarity of this star can be explained with inhomogeneous mixing, where it formed out of an environment rich in SNe Type Ia yields compared to SNe II yields, for its [Fe/H].

       \begin{figure}
   \centering
   \includegraphics[width=\hsize-1.cm]{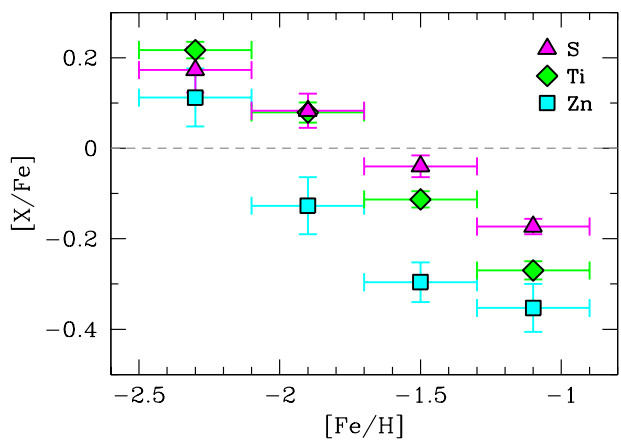}
      \caption{Average abundance ratios of: S (magenta triangles), Ti (green diamonds), and Zn (cyan squares), over Fe in four metallicity bins, from our FLAMES data and \citet{Skuladottir2015b}. The x-errorbars show the size of the [Fe/H] bin, and the y-errorbars are the error of the mean. 
      }
         \label{fig:join}
   \end{figure}

Despite the large scatter in [Zn/Fe] values at a given [Fe/H], our data reveal a clear downwards trend of [Zn/Fe] with [Fe/H], see Fig.~\ref{fig:ZnFe}, similar to that seen in $\alpha$-elements. This is shown more clearly in Fig.~\ref{fig:join}, where the average measured values of S, Ti and Zn over Fe are shown for four [Fe/H] bins. The errorbars of these mean values are defined as $\delta_\text{avg}=\sigma/\sqrt{(N-1)}$, where $\sigma$ is the standard deviation of the scatter, and $N$ is the number of stars in each bin. Stars with only an upper limit are excluded. In the case of Zn, stars with only upper limits all have low quality spectra, $11 \leq \text{S/N}_\text{blue}\leq18$, while the average of the sample is $<\text{S/N}_\text{blue}>\approx 24$ with a standard deviation of $\sigma_\text{S/N}\approx6$. Therefore, we can assume that the average value of [Zn/Fe] for the stars with only upper limits is not significantly different from the rest of the sample, and can thus be safely excluded in Fig.~\ref{fig:join}.

Finally we note that the two most metal-rich bins are uniquely consistent with having the same [Zn/Fe], within the errors. One of the production-sites of Zn is neutron-capture processes, which are more effective in massive Type II supernovae at higher metallicities \citep{Nomoto2013}. This metallicity dependence of the yields could therefore be a possible explanation for this apparent saturation in the abundance ratios.

\begin{figure}
   \centering
   \includegraphics[width=\hsize-1.cm]{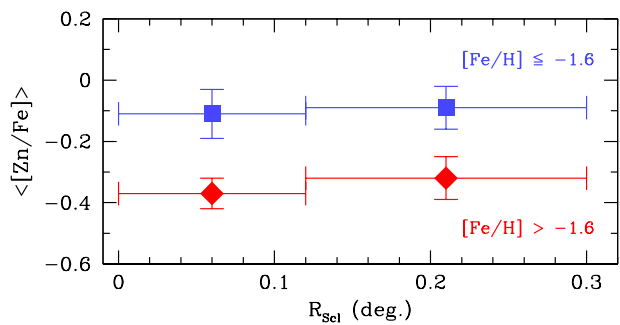}
      \caption{Average abundance ratios of [Zn/Fe] as a function of distance from the center of Sculptor, $R_\textsl{Scl}$, separated by metallicity. Blue squares are stars with $\text{[Fe/H]}\leq-1.6$ and red diamonds have $\text{[Fe/H]}>-1.6$. The x-errorbars show the size of the bins, and y-errorbars the error of the mean.
      }
         \label{fig:ellrad}
   \end{figure}

Recently \citealt{Duffau2017} found decreasing [Zn/Fe] in the Milky Way with smaller distances to the Galactic center. To test for evidence of spatial variation in Sculptor, we divided our sample into an outer ($R_\textsl{Scl}>0.12$ deg.) and inner region  ($R_\textsl{Scl}\leq0.12$ deg.)  sub-samples, which contain similar number of stars ($\approx40$), see Fig.~\ref{fig:ellrad}. The metallicity gradient in our sample is minimal, and the difference in $<\text{[Fe/H]}>$ is $\leq0.03$~dex between the inner and outer regions, both for the high- and low-metallicity groups. In our data there is no significant difference in [Zn/Fe] between the inner and outer regions. Considering the errors and that all the stars are within 18 arcminutes of the centre of Sculptor this is perhaps unsurprising. However, we do note that the dispersion in [Zn/Fe] is smallest for the metal-rich stars in the inner region, $\sigma \approx0.2$, while it is $\sigma \approx0.3$ in the other three sub-samples. 

Compared to other iron-peak elements, the evolution of Zn in Sculptor seems rather complicated, with a decreasing trend of [Zn/Fe] with [Fe/H], similar to $\alpha$-elements. But the abundance ratio also has a significant scatter at a given [Fe/H], and a possible saturation of the trend at the highest metallicities in Sculptor.

\subsection{Correlation of the Zn scatter with other elements}

The [Zn/Fe] and [$\alpha$/Fe] abundance ratios in Sculptor show a similar trend with [Fe/H] and are therefore correlated, see Fig. \ref{fig:join}. This results from these elements not primarily being produced in Type Ia SN, rather than necessarily suggesting a common production channel. If, on the other hand, the abundance scatter is correlated with that of another element, further clues on the nucleosynthesis of Zn may be drawn.

To test this, for each star we calculate the deviation from the mean, $\Delta{\text{[X/Fe]}}$, at a given [Fe/H]. Without the effects of the mean trend, we can check for correlations between the scatter of Zn, $\Delta\text{[Zn/Fe]}$, and the scatter in other elements, $\Delta\text{[X/Fe]}$. A significant number of elements have previously been measured for the sample stars (\citealt{Tolstoy2009}; \citealt{Skuladottir2015b}; Skúladóttir 2016; Hill et al. in prep.) and we also include measurements from the literature (\citealt{Shetrone2003,Geisler2005}; Hill et al. in prep.). Stars with only upper limits for Zn or the relevant reference element were excluded, as were stars with errors $\delta_\text{[Zn/Fe]}>0.50$.

       \begin{figure}
   \centering
   \includegraphics[width=\hsize-2.2cm]{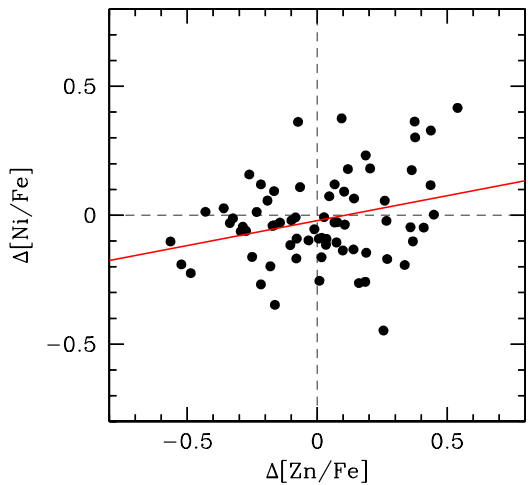}
      \caption{Scatter of Ni, $\Delta\text{[Ni/Fe]}$, as a function of the scatter of Zn, $\Delta\text{[Ni/Fe]}$. Red line shows best fit through the measurements.
      }
         \label{fig:correlation}
   \end{figure}

The following elements had sufficient measurements available for this comparison: O, Na, Mg, Si, Ca, Ti, Sc, Cr, Mn, Fe II, Co, Ni, Ba, La and Nd. In each case, $>50$ stars had measurements of these elements and Zn, with the exception of O ($\approx$20 stars) and Sc ($\approx$30 stars). A statistically significant correlation with the scatter of Zn was found only for one element: Ni. A comparison with $n=70$ stars gave a correlation coefficient of $r=0.29$ and a $t$-value of $t=2.5$, while the correlation with all other elements have t<1.2.

Although the correlation is significant, it can either be astrophysical or arise from some aspect of our analysis. The Ni abundance measurements were made using equivalent width measurements, with another set of spectra (Hill et al. in prep.), so systematics in the analysis, such as uncertainties in the continuum placement can be excluded. However, as the same stellar parameters are used, if errors of $T_\textsl{eff}$, $\log g$ and $v_t$ affect the Zn and Ni lines in a similar fashion, then this co-variance could produce a correlation in the way seen here.

To test if this is a plausible explanation, 70 artificial stars were given two Gaussian random errors, $\delta_\text{Zn}$ and $\delta_\text{Ni}$, so that the average amplitude of these errors in the sample would be equal to the average measurement errors of Zn and Ni, respectively. Typically, the errors of any two elements in a sample of stars can be expected to be correlated, since on average brighter stars have lower errorbars on all elements compared to fainter stars. However, Zn and Ni were measured with two different sets of spectra, so there is no significant correlation between the measured errorbars ($r_\textsl{err}\lesssim0.05$). Therefore, no correlation was added for the artificial stars. 

In addition to the artificial measurement errors, a Gaussian error due to stellar parameters, $\delta_{\rm sp}$, was added; the same value was assumed for both elements, with an average amplitude of 0.04 dex (the average stellar parameter error for [Zn/Fe]):
\begin{eqnarray}
\delta\text{[Zn/Fe]}_\textsl{artificial}=\delta_\text{Zn} +\delta_\textsl{sp}\\
\delta\text{[Ni/Fe]}_\textsl{artificial}=\delta_\text{Ni} +\delta_\textsl{sp}
\end{eqnarray}

This was done to see if the errors of the stellar parameters were enough to create this correlation, in case they happened to work exactly the same for the Zn and Ni lines. Doing $10^6$ realizations of this simple exercise revealed that $r\geq0.29$ in $<3\%$ of the cases. Even when the average amplitude of $\delta_\textsl{sp}$ was changed to 0.10 dex (which is unreasonably high), $r<0.29$ in $\approx 90\%$ of cases. The errors on the stellar parameters are therefore an unlikely source of the correlation between the scatter observed in 
[Ni/Fe] and [Zn/Fe] when plotted against [Fe/H]. 

Although the validity of this correlation is not unequivocal, see Fig.~\ref{fig:correlation}, we do note that with an atomic number of $Z=28$, Ni is the next even element to Zn, $Z=30$. We therefore conclude that it is not improbable that this correlation is physical and comes from the co-production of these elements, although this possibility needs to be confirmed with data of higher quality.

\setcounter{table}{7}
\begin{table}
\caption{Measured chemical abundance ratios in Sculptor, [Fe/H] and [X/Fe] (same as in Fig.~\ref{fig:join}), used as initial and final conditions, and SN Type Ia nucleosynthetic yields, $\mathcal{Y}_\text{X}$, i.e. synthesized mass of each element X in one SN Type Ia \citep{Iwamoto1999}.}
\label{tab:yields} 
\centering
\begin{tabular}{c c c c}
\hline\hline
Ele. &	Initial cond.$^\text{1}$	 &	Final cond.$^\text{1}$   & $\mathcal{Y}_{\text{X}}$	\\
		X			    &   [X/Fe]$^{\textsl{in}}$                  & [X/Fe]$^{\textsl{fin}}$                &   $M_\odot$\\
\hline
Fe		&		$-2.3$							&		$-1.5$								&		$(7.6\pm0.3)\cdot10^{-1}$		\\ 
Zn 	&		$+0.11 \pm 0.06$		& 		$-0.30 \pm 0.04$			& 		$(7.1\pm2.2)\cdot10^{-5}$		\\
Ti 	&		$+0.22 \pm 0.02$		& 		$-0.11 \pm 0.02$			& 		$(7.7\pm1.2)\cdot10^{-4}$		\\
S 		&		$+0.17 \pm 0.06$		& 		$-0.04 \pm 0.02$			& 		$(1.2\pm0.1)\cdot10^{-1}$		\\
\hline
\multicolumn{4}{l}{$^1$[Fe/H] for Fe}\\
\end{tabular}
\end{table}

\section{Zn evolution: The influence of Type Ia SNe}

The relevant contributions of SN Type II and Type Ia, and their different time scales, are the main drivers of decreasing [$\alpha$/Fe] with metallicity. The $\alpha$-like behaviour of Zn, as shown in Fig.~\ref{fig:join}, has previously also been observed in the Milky Way (\citealt{NissenSchuster2011,Barbuy2015,Duffau2017}), and seems to suggest a similar explanation for the evolution of the [Zn/Fe] ratio.

Theoretical yields are also broadly consistent with this, where the mass fraction in the ejecta, $M_\text{Zn}/M_\text{Fe}$, is significantly higher from SN Type~II compared to Type~Ia (e.g. \citealt{Iwamoto1999}). To test this more quantitatively, we compare the measured chemical abundances in Sculptor (see Table~\ref{tab:yields}) with simple estimates using theoretical SN yields. We consider the change from an initial metallicity of $\text{[Fe/H]}^\textsl{in}=-2.3$ to the final value of $\text{[Fe/H]}^\textsl{fin}=-1.5$, where $\text{[X/Fe]}^\textsl{in/fin}$ are averaged chemical abundances in Sculptor, over $\pm$0.2 in [Fe/H] in each case, see Fig.~\ref{fig:join} and Table~\ref{tab:yields}.

The initial conditions, at $\text{[Fe/H]}^\textsl{in}=-2.3$, can be described with the following equations:
\begin{eqnarray}
\text{[Fe/H]}^\textsl{in}= \log \left( \frac{M_\text{Fe}^\textsl{in}}{M_\text{g}^\textsl{in}} \right)-\log \left( \frac{M_\text{Fe}}{M_\text{g}} \right)_\odot \\
\text{[X/Fe]}^\textsl{in}= \log \left( \frac{M_\text{X}^\textsl{in}}{M_\text{Fe}^\textsl{in}} \right)-\log \left( \frac{M_\text{X}}{M_\text{Fe}} \right)_\odot
\end{eqnarray}
We assume the initial mass of gas to be $M_\textsl{g}^\textsl{in}=10^7M_\odot$ \citep{Salvadori2015}, but the final result is independent of this value. Combining these two equations we get the initial elemental mass $M_{X}^\textsl{in}$ for each element. For simplicity, we assume that SNe Type Ia dominate the chemical evolution from $\text{[Fe/H]}=-2.3$ to $-1.5$, so the mass created of each element is $\approx M_\text{X}^\text{Ia}$. The final conditions are then:

\begin{eqnarray}
\text{[X/Fe]}^\textsl{fin}&=& \log \left( \frac{M_\text{X}^\textsl{in}+ M_\text{X}^\textsl{Ia} }{M_\text{Fe}^\textsl{in}+ M_\text{Fe}^\text{Ia} } \right)-\log \left( \frac{M_\text{X}}{M_\text{Fe}} \right)_\odot \nonumber\\
&=& \log \left( \frac{M_\text{X}^\textsl{in}+ \mathcal{Y}_\text{X}  N^\textsl{Ia} }{M_\text{Fe}^\textsl{in}+ \mathcal{Y}_\text{Fe} N^\textsl{Ia} } \right)-\log \left( \frac{M_\text{X}}{M_\text{Fe}} \right)_\odot \label{eq:Xout}
\end{eqnarray}
Here $\mathcal{Y}_X$ is the mass ejected for element X in one SN Type~Ia, which is obtained by averaging over all models in \citet{Iwamoto1999}, see Table~\ref{tab:yields}. The listed uncertainty on $\mathcal{Y}_\text{X}$ is the error of the mean between different models. With the theoretical values of  $\mathcal{Y}_X$, we are able to calculate how many Type~Ia SN are needed, $N^\textsl{Ia}$, to obtain the final value of $\text{[Zn/Fe]}^\textsl{fin}=-0.29$. Testing whether this scenario is also consistent with the measured values of $\text{[Ti/Fe]}^\textsl{fin}$ and $\text{[S/Fe]}^\textsl{fin}$, we use the value of $N^\textsl{Ia}$ with equation (\ref{eq:Xout}) to self-consistently calculate:

\begin{eqnarray}
\text{[Ti/Fe]}^{\textsl{fin}}_\textsl{calc} &=&-0.08\pm 0.02 \\
\text{[S/Fe]}^{\textsl{fin}}_\textsl{calc}  &=&-0.04\pm 0.02 
\end{eqnarray}
The result is consistent with the measured values, although the uncertainties here only include the propagated error of the mean of $\mathcal{Y}_\text{X}$ (as listed Table~\ref{tab:yields}). This is a simple estimate and for a more realistic model other effects have to be accounted for, e.g. the contribution of Type~II SN, PISN, inflows and outflows of gas and inhomogeneous metal mixing. Our estimates are, however, sufficient to show that SN Type Ia yields can self-consistently explain the observed mean trends of Zn, Ti and S with Fe, in the Sculptor dSph.

\section{Zinc in the Local Group and beyond}

\begin{figure*}
   \centering
   \includegraphics[width=\hsize-1.5cm]{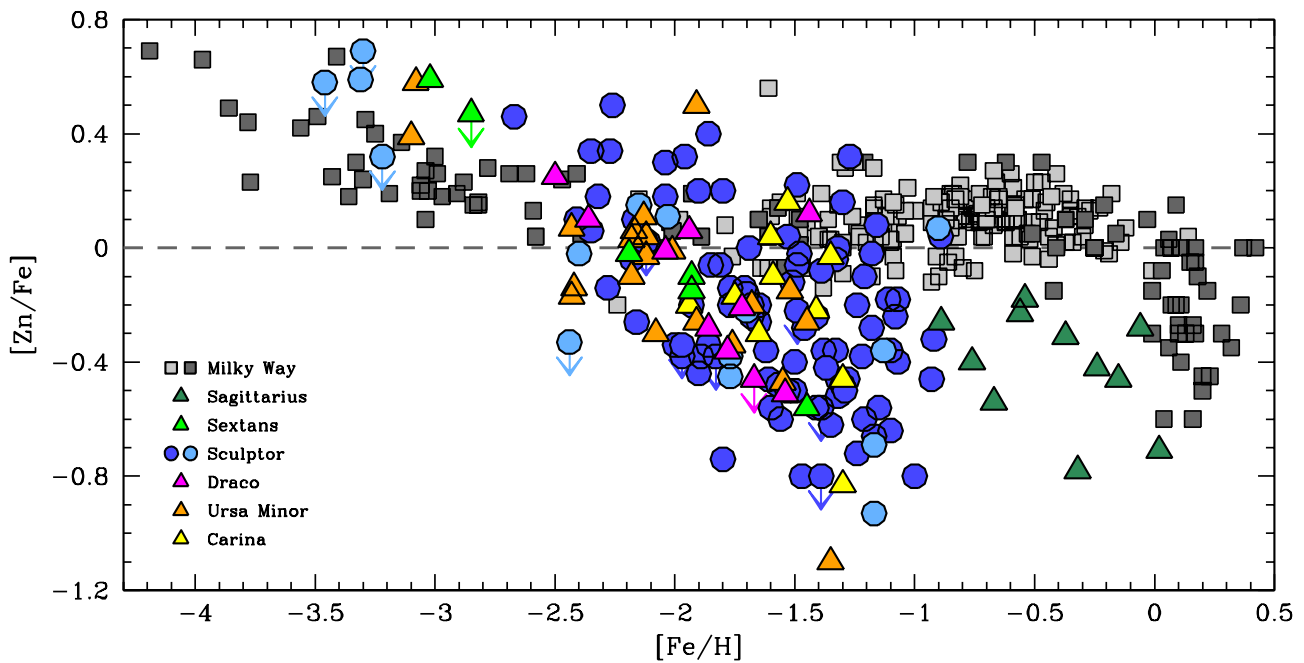}
      \caption{Measurements of [Zn/Fe] in individual stars in the Local Group. Upper limits larger than $\text{[Zn/Fe]}=0.7$ are excluded from the plot, as are stars with errors $\delta_\text{[Zn/Fe]}>0.5$. Blue circles show stars from this work on the Sculptor dSph, and light blue circles are Sculptor stars from the literature. Triangles are stars from other dwarf galaxies; included are (from highest to lowest stellar mass according to \citealt{McConnachie2012}): Sagittarius stars are dark green, Sextans is green, Draco is magenta, Ursa Minor is orange, and yellow is for Carina stars. Squares are Milky way stars, where giant stars are dark gray and dwarf stars light gray. References (Number of stars): \textit{Sagittarius:} \citealt{Sbordone2007} (11). \textit{Sextans:} \citealt{Shetrone2001} (5); \citealt{Honda2011} (1). \textit{Sculptor:} \citealt{Shetrone2003} (1); \citealt{Geisler2005} (1); \citealt{Kirby2012} (1); \citealt{Jablonka2015} (4 upper limits); \citealt{Skuladottir2015a} (1); \citealt{Simon2015} (1); Hill et al. in prep (6).  \textit{Draco:} \citealt{Shetrone2001} (5); \citealt{Cohen2009} (5). \textit{Ursa Minor:} \citealt{Shetrone2001} (6); \citealt{Sadakane2004} (3); \citealt{Cohen2010} (10); \citealt{Ural2015} (2). \textit{Carina:} \citealt{Shetrone2003} (5); \citealt{Venn2012} (5). \textit{Milky Way:} \citealt{Reddy2003,Reddy2006};  \citealt{Cayrel2004}; \citealt{NissenSchuster2011}; \citealt{Ishigaki2013}; \citealt{Bensby2014} (only including stars with errors $\delta_\text{[Zn/Fe]}\leq0.2$); \citealt{Barbuy2015}.
      }
         \label{fig:ZnFe2}
   \end{figure*}

At present, stellar abundances of Zn have only been measured in a handful of galaxies in the Local Group, for a limited number of stars. We have collected the available literature data in Fig.~\ref{fig:ZnFe2}, most of which can also be found in the compilation of \citet{Berg2015}. The [Zn/Fe] measurements in dSph galaxies are not many and only cover a restricted number of stars and [Fe/H] range in each system. This limits the conclusions regarding the entire chemical enrichment history of Zn in these galaxies. Overall the currently available observations show similar trends in dSph galaxies to Sculptor, mainly low [Zn/Fe] values with an indication of spread and/or outliers.

Similar to Sculptor, abundance measurements of stars in the main body of the Sagittarius dSph show low values of [Zn/Fe] \citep{Sbordone2007}, although in this significantly larger galaxy the measured stars are at a higher [Fe/H]. Some scatter is present in the [Zn/Fe] abundance ratio in Sagittarius; however, no error estimates were published for these results, so it is possible that the spread is within what is expected from measurement uncertainties. In this metallicity range, $-1\lesssim\text{[Fe/H]}\lesssim 0$, the measurements of $\alpha$-elements also show a declining trend in Sagittarius \citep{Sbordone2007}. This is consistent with the explanation that the low [Zn/Fe] arise from an increased contribution from SNe Type Ia.

In the Carina dSph a decreasing trend with [Fe/H] due to SNe Type Ia contribution is not obvious in the available measurements of $\alpha$-elements \citep{Venn2012,Lemasle2012,Fabrizio2015}, yet low [Zn/Fe] abundances are observed, see Fig.~\ref{fig:ZnFe2}. We note that the star Venn612 in Carina with $\text{[Zn/Fe]}=-0.8$ (see Fig.~\ref{fig:ZnFe2}) has an abnormal chemical abundance pattern with low values of the $\alpha$-elements in general, and some of the iron-peak elements, compared to other stars in Carina \citep{Venn2012}. \citet{Venn2012} concluded that this arises from inhomogeneous mixing, where this star was formed in a pocket of interstellar medium rich in SNe Type Ia ejecta, relative to SNe Type II. A similar interpretation has been applied to the star ET0381 at $\text{[Fe/H]}=-2.4$ from Sculptor from \citet{Jablonka2015}, implying that inhomogeneous mixing is commonplace in the earliest stages of dwarf galaxy evolution. If this outlier in Carina, Venn612, is excluded, the other measurements agree within the errorbars, with subsolar mean value of $\text{[Zn/Fe]}=-0.14\pm0.07$, where the error here is $\delta_\text{avg}$. 

In the dSph galaxies Sextans, Draco and Ursa Minor, [Zn/Fe] shows a decrease with [Fe/H], like in Sculptor. Similar to Venn612, the extremely Zn-poor star in Ursa minor, UMi-COS171 at $\text{[Zn/Fe]}=-1.1$ \citep{Cohen2010}, also has low abundance ratios in most other measured chemical elements with respect to Fe, in particular Mg, Sc, Ti, Mn and Ni. Its overall abundance pattern is thus unusual for Ursa Minor. Both in Draco and Ursa Minor the measurements show evidence of some scatter in [Zn/Fe]. There is no evidence for scatter in [Zn/Fe] abundance ratios in Sextans, but with only six measurements, it is impossible to conclude whether scatter is present in this galaxy. 

Usually, only one line is used when Zn is measured in dwarf galaxy stars, and therefore errorbars are not always specified. Thus it is difficult to conclude overall how significant the measured spread in [Zn/Fe] is from previous work. However, all the Zn abundances in the literature are derived with a spectra of higher resolution than this work, and so on average the errorbars are expected to be smaller.

At the lowest metallicities, $\text{[Fe/H]}\lesssim-2$, the abundance ratios of [Zn/Fe] in Sculptor and the Milky Way halo are consistent, see Fig.~\ref{fig:ZnFe2}. But the trend of [Zn/Fe] at higher [Fe/H] in the Milky Way is quite complicated. Typical $\alpha$-elements in the Milky Way disc, show a decline in [$\alpha$/Fe] in the range $-1\leq\text{[Fe/H]}\leq0$ (e.g. \citealt{Chen2002,Venn2004a}). This is not seen in [Zn/Fe], which is broadly flat. However, there is structure in this flatness as the thick disc has higher [Zn/Fe] than the thin disc, similar to [$\alpha$/Fe] (e.g. \citealt{Reddy2006,Mikolaitis2017}). Recent work by \citet{Duffau2017} shows different [Zn/Fe] abundant ratios in dwarf and giant stars, which they attribute to the giants being more confined to the inner Galactic thin disc. So from these works, it is clear that the Galactic disc underwent a complex enrichment in Zn: the thick disc is different to the thin disc, and the inner and outer thin disc also differ in Zn. In the Milky Way bulge, at $\text{[Fe/H]}\geq -0.1$, there is a clear decline in [Zn/Fe], and a significant scatter in giant stars. \citet{Barbuy2015} explained this with SN Type Ia contribution. This is not confirmed by the microlensed dwarf and subgiant stellar sample of \citet{Bensby2013} in the bulge, and this discrepancy is currently unresolved. However, in the metallicity range of Sculptor, Milky Way giant and dwarf stars are in agreement.

 Comparing the Zn evolution of Sculptor to the Milky Way, the decrease of [Zn/Fe] and sub-solar values at the highest metallicities seems a logical result from more contribution from SN Type Ia. So to explain the lack of low [Zn/Fe] in the dwarf stars of \citet{Bensby2013}, other sources of Zn have to be included. The metallicity dependence of Zn yields in SN Type II \citep{Nomoto2013} is expected to have more influence in the Milky Way, compared to Sculptor, where there has been no star formation over the last $\approx6$~Gyr. In addition, theoretical calculations have suggested that a significant proportion of Zn in Milky Way stars has an origin in  hypernovae, both at low and high metallicities \citep{UmedaNomoto2002,Kobayashi2006,Barbuy2015}. Whether these high energy supernovae play an important role in the chemical evolution histories in the much smaller dwarf galaxies is still not understood. 

Due to its volatile nature, Zn is not significantly depleted onto dust, and can therefore be accurately measured in the interstellar medium of damped Lyman-alpha systems (DLAs) \citep{SpitzerJenkins1975,Savage1996,Jenkins2009,Vladilo2011}. Since it belongs to the upper group of iron-peak elements, it has often been used as a proxy for Fe. Although this might be reasonable in some cases, caution is advised, since it is clear from measurements of stellar abundances in the Local Group that the behaviour of [Zn/Fe] with [Fe/H] is complicated, and environment dependent. A more detailed comparison between chemical abundances in DLAs and dwarf galaxies will be done in a following paper: Skúladóttir et al. in prep.

%
\section{Conclusions}

A sample of $\approx$100 RGB stars in Sculptor was observed with VLT FLAMES/GIRAFFE, using the HR7A setting, which covers the wavelength range $\sim$4700-4970~\AA. The sample consists of 15~new stars, and 86~stars which have previously been observed in other wavelength regions with VLT/FLAMES (\citealt{Tolstoy2009,Skuladottir2015b}, Hill et al. in prep.). The Fe and Ti measured from our HR7A spectra generally agree well with previously reported values from other observations, showing a decreasing trend of [Ti/Fe] with [Fe/H].

The [Zn/Fe] ratios in Sculptor decline with increasing [Fe/H]. This can self-consistently be explained by an increasing contribution of SNe Type Ia yields to the environment; such yields contain large amounts of Fe compared to Zn and the $\alpha$-elements ([Zn/Fe]<0, [$\alpha$/Fe]<0). No spatial variation of [Zn/Fe] is observed, unsurprisingly since the entire sample is centrally concentrated.

Stars in Sculptor have a large range of Zn abundances, $-0.9\lesssim\text{[Zn/Fe]}\lesssim+0.5$. The scatter is quite significant, and cannot be explained solely by the measurement errors. However, with the current data, it is unclear whether there is an underlying uniform scatter, or if the sample shows a narrow trend with a few outliers. 

The measured scatter of [Zn/Fe] shows a statistically significant correlation with the scatter in [Ni/Fe]. An artificial correlation due to our abundance analysis has been shown to be unlikely, but better data are needed to confirm the validity of this correlation.

The results presented here have considerably increased the measurements of Zn abundances in Sculptor, and are currently by far the largest sample of Zn measurements in any dwarf galaxy. Our findings are broadly consistent with the sparse abundance measurements in other dSph galaxies, which also show sub-solar values of [Zn/Fe] and indications of scatter. However, the results differ from the measurements of the different component in the Milky Way.
Taken together, observational evidence in Local Group galaxies makes it clear that [Zn/Fe] is not constant in different environments, nor over different [Fe/H] scales.

\begin{acknowledgements}
The authors are indebted to the International Space Science Institute (ISSI), Bern, Switzerland, for supporting and funding the international team `First stars in dwarf galaxies'. \'A.S. thanks Ken Nomoto and Karin Lind for useful advice and insightful suggestions. \'A.S. acknowledges funds from the Alexander von Humboldt Foundation in the framework of the Sofja Kovalevskaja Award endowed by the Federal Ministry of Education and Research. E.T. gratefully acknowledges support to visit to the IoA from the Sackler Fund for
Astronomy. S.S. was supported by the European Commission through a Marie-Curie Fellowship, project PRIMORDIAL, grant nr. 700907, and by the Italian Ministry of Education, University, and Research (MIUR) through a Rita Levi Montalcini Fellowship. M.P. is grateful to the Kapteyn Astronomical Institute for their generous hospitality during multiple visits while this work was in progress.
\end{acknowledgements}

\bibliography{heimildir}

\clearpage
\pagebreak

\setcounter{table}{1}
\begin{sidewaystable*}
\label{tab4:abundances}   
\caption{Stellar parameters, abundances and S/N.}
\tiny
\centering
\tabcolsep=0.08cm   
\begin{tabular}{lccccccrcccrc}     
\hline\hline       
Star	&	[Fe/H]	&	$\delta_\text{[Fe/H]}$	&	S/N$_\text{blue}$	&	S/N$_\text{red}$	&	$N_\text{Ti}$	&	$\delta_\text{noise}\text{(Ti)}$	&	[Ti/Fe]	&	$\delta_\text{[Ti/Fe]}$	&	$N_\text{Zn}$	&	$\delta_\text{noise}\text{(Zn)}$	&	[Zn/Fe]	&	$\delta_\text{[Zn/Fe]}$\\
\hline
ET0024	&$	-1.24	$&	0.10	&	23	&	36	&	22	&	0.08	&$	-0.09	$&	0.16	&	1	&	0.40	&$	-0.20	$&	0.41	\\
ET0026	&$	-1.80	$&	0.16	&	34	&	47	&	19	&	0.07	&$	-0.04	$&	0.09	&	1	&	0.34	&$	-0.74	$&	0.36	\\
ET0027	&$	-1.50	$&	0.13	&	31	&	53	&	27	&	0.06	&$	-0.23	$&	0.10	&	1	&	0.16	&$	-0.40	$&	0.21	\\
ET0028	&$	-1.22	$&	0.11	&	31	&	39	&	25	&	0.05	&$	-0.39	$&	0.11	&	1	&	0.38	&$	-0.38	$&	0.39	\\
ET0031	&$	-1.68	$&	0.17	&	29	&	51	&	19	&	0.06	&$	-0.03	$&	0.09	&	1	&	0.18	&$	-0.24	$&	0.23	\\
ET0033	&$	-1.77	$&	0.16	&	24	&	40	&	19	&	0.05	&$	-0.11	$&	0.13	&	1	&	0.24	&$	-0.14	$&	0.29	\\
ET0043	&$	-1.24	$&	0.16	&	22	&	34	&	20	&	0.07	&$	-0.24	$&	0.13	&	1	&	0.46	&$	-0.72	$&	0.48	\\
ET0048	&$	-1.90	$&	0.19	&	46	&	71	&	14	&	0.07	&$	0.14	$&	0.10	&	1	&	0.20	&$	0.20	$&	0.25	\\
ET0051	&$	-0.92	$&	0.12	&	31	&	47	&	23	&	0.07	&$	-0.36	$&	0.17	&	1	&	0.34	&$	-0.32	$&	0.35	\\
ET0054	&$	-1.81	$&	0.16	&	31	&	56	&	20	&	0.07	&$	0.06	$&	0.09	&	1	&	0.20	&$	-0.06	$&	0.25	\\
ET0057	&$	-1.33	$&	0.13	&	36	&	48	&	24	&	0.05	&$	-0.15	$&	0.10	&	1	&	0.22	&$	-0.36	$&	0.24	\\
ET0059	&$	-1.53	$&	0.16	&	34	&	50	&	24	&	0.07	&$	-0.25	$&	0.11	&	1	&	0.32	&$	-0.50	$&	0.34	\\
ET0060	&$	-1.56	$&	0.15	&	28	&	46	&	12	&	0.10	&$	-0.14	$&	0.12	&	1	&	0.18	&$	-0.60	$&	0.21	\\
ET0062	&$	-2.27	$&	0.18	&	18	&	33	&	11	&	0.13	&$	0.24	$&	0.17	&	1	&	0.36	&$	0.34	$&	0.41	\\
ET0063	&$	-1.18	$&	0.19	&	26	&	41	&	22	&	0.05	&$	-0.38	$&	0.13	&	1	&	0.28	&$	-0.02	$&	0.30	\\
ET0064	&$	-1.38	$&	0.14	&	34	&	41	&	20	&	0.06	&$	-0.10	$&	0.11	&	1	&	0.24	&$	-0.36	$&	0.26	\\
ET0066	&$	-1.30	$&	0.14	&	24	&	38	&	20	&	0.08	&$	-0.23	$&	0.11	&	1	&	0.48	&$	-0.48	$&	0.49	\\
ET0067	&$	-1.65	$&	0.16	&	26	&	40	&	23	&	0.08	&$	0.13	$&	0.09	&	1	&	0.22	&$	-0.26	$&	0.26	\\
ET0069	&$	-2.11	$&	0.20	&	28	&	47	&	11	&	0.08	&$	0.20	$&	0.13	&	1	&	0.34	&$	0.02	$&	0.37	\\
ET0071	&$	-1.35	$&	0.14	&	30	&	48	&	24	&	0.08	&$	-0.16	$&	0.11	&	1	&	0.32	&$	-0.62	$&	0.34	\\
ET0073	&$	-1.53	$&	0.17	&	27	&	39	&	23	&	0.05	&$	-0.07	$&	0.09	&	1	&	0.38	&$	0.04	$&	0.41	\\
ET0083	&$	-1.97	$&	0.18	&	29	&	49	&	17	&	0.07	&$	0.17	$&	0.10	&	1	&	0.30	&$	-0.38	$&	0.34	\\
ET0094	&$	-1.86	$&	0.15	&	16	&	27	&	17	&	0.08	&$	0.07	$&	0.11	&	1	&	0.46	&$	-0.34	$&	0.49	\\
ET0095	&$	-2.16	$&	0.20	&	30	&	44	&	13	&	0.08	&$	0.16	$&	0.11	&	1	&	0.26	&$	-0.26	$&	0.31	\\
ET0103	&$	-1.21	$&	0.13	&	18	&	29	&	18	&	0.07	&$	-0.27	$&	0.11	&	1	&	0.34	&$	-0.10	$&	0.36	\\
ET0104	&$	-1.62	$&	0.17	&	22	&	35	&	16	&	0.08	&$	0.02	$&	0.11	&	1	&	0.38	&$	-0.36	$&	0.41	\\
ET0109	&$	-1.85	$&	0.11	&	30	&	48	&	30	&	0.07	&$	0.02	$&	0.13	&	1	&	0.26	&$	-0.06	$&	0.29	\\
ET0112	&$	-2.04	$&	0.14	&	37	&	56	&	19	&	0.07	&$	0.07	$&	0.09	&	1	&	0.12	&$	0.18	$&	0.17	\\
ET0113	&$	-2.18	$&	0.19	&	37	&	58	&	17	&	0.07	&$	0.17	$&	0.10	&	1	&	0.18	&$	-0.04	$&	0.24	\\
ET0121	&$	-2.35	$&	0.20	&	22	&	43	&	10	&	0.09	&$	0.17	$&	0.16	&	1	&	0.20	&$	0.34	$&	0.25	\\
ET0126	&$	-1.11	$&	0.16	&	28	&	43	&	21	&	0.06	&$	-0.37	$&	0.11	&	1	&	0.34	&$	-0.18	$&	0.36	\\
ET0132	&$	-1.50	$&	0.15	&	29	&	47	&	22	&	0.06	&$	0.05	$&	0.09	&	1	&	0.22	&$	-0.50	$&	0.26	\\
ET0133	&$	-1.07	$&	0.15	&	26	&	39	&	22	&	0.06	&$	-0.20	$&	0.14	&	1	&	0.26	&$	-0.40	$&	0.28	\\
ET0137	&$	-0.89	$&	0.18	&	28	&	46	&	18	&	0.07	&$	-0.56	$&	0.19	&	1	&	0.28	&$	0.04	$&	0.29	\\
ET0138	&$	-1.70	$&	0.15	&	24	&	42	&	16	&	0.07	&$	-0.14	$&	0.10	&	1	&	0.24	&$	-0.16	$&	0.28	\\
ET0139	&$	-1.41	$&	0.11	&	29	&	37	&	17	&	0.08	&$	-0.28	$&	0.13	&	1	&	0.22	&$	-0.56	$&	0.25	\\
ET0141	&$	-1.68	$&	0.15	&	28	&	48	&	19	&	0.08	&$	-0.04	$&	0.10	&	1	&	0.24	&$	-0.20	$&	0.28	\\
ET0145	&$	-1.51	$&	0.14	&	34	&	47	&	13	&	0.09	&$	-0.31	$&	0.10	&	1	&	0.22	&$	-0.12	$&	0.25	\\
ET0150	&$	-0.93	$&	0.11	&	22	&	34	&	13	&	0.08	&$	-0.45	$&	0.16	&	1	&	0.36	&$	-0.46	$&	0.37	\\
ET0151	&$	-1.77	$&	0.16	&	25	&	42	&	18	&	0.08	&$	0.11	$&	0.10	&	1	&	0.22	&$	-0.20	$&	0.26	\\
ET0158	&$	-1.80	$&	0.21	&	27	&	45	&	12	&	0.07	&$	0.08	$&	0.10	&	1	&	0.36	&$	0.20	$&	0.39	\\
ET0160	&$	-1.16	$&	0.14	&	26	&	36	&	13	&	0.09	&$	-0.28	$&	0.13	&	1	&	0.52	&$	-0.50	$&	0.54	\\
ET0163	&$	-1.86	$&	0.21	&	25	&	38	&	12	&	0.13	&$	-0.01	$&	0.16	&	1	&	0.20	&$	0.40	$&	0.26	\\
ET0164	&$	-1.89	$&	0.22	&	24	&	42	&	17	&	0.05	&$	0.11	$&	0.11	&	1	&	0.20	&$	-0.38	$&	0.25	\\
ET0165	&$	-1.10	$&	0.17	&	25	&	38	&	19	&	0.09	&$	-0.31	$&	0.13	&	1	&	0.22	&$	-0.36	$&	0.25	\\
ET0166	&$	-1.49	$&	0.15	&	21	&	35	&	16	&	0.07	&$	0.02	$&	0.09	&	1	&	0.44	&$	-0.06	$&	0.46	\\
ET0168	&$	-1.10	$&	0.17	&	18	&	31	&	16	&	0.09	&$	-0.29	$&	0.13	&	1	&	0.38	&$	-0.64	$&	0.40	\\
ET0173	&$	-1.47	$&	0.10	&	20	&	34	&	17	&	0.10	&$	-0.31	$&	0.15	&	1	&	0.20	&$	-0.80	$&	0.22	\\
ET0198	&$	-1.16	$&	0.17	&	22	&	31	&	15	&	0.09	&$	-0.32	$&	0.11	&	1	&	0.22	&$	0.08	$&	0.26	\\
ET0200	&$	-1.49	$&	0.19	&	23	&	39	&	15	&	0.08	&$	-0.07	$&	0.12	&	1	&	0.38	&$	0.22	$&	0.42	\\
ET0202	&$	-1.32	$&	0.19	&	21	&	35	&	19	&	0.08	&$	-0.21	$&	0.11	&	1	&	0.38	&$	-0.52	$&	0.40	\\

\hline
\end{tabular}
\end{sidewaystable*}

\setcounter{table}{1}
\begin{sidewaystable*}
\label{tab4:abundances2}
\tiny
\centering
\tabcolsep=0.08cm
\caption{Stellar parameters, abundances and S/N, continued.}   
\begin{tabular}{lccccccrcccrc}       
\hline\hline       
Star	&	[Fe/H]	&	$\delta_\text{[Fe/H]}$	&	S/N$_\text{blue}$	&	S/N$_\text{red}$	&	$N_\text{Ti}$	&	$\delta_\text{noise}\text{(Ti)}$	&	[Ti/Fe]	&	$\delta_\text{[Ti/Fe]}$	&	$N_\text{Zn}$	&	$\delta_\text{noise}\text{(Zn)}$	&	[Zn/Fe]	&	$\delta_\text{[Zn/Fe]}$\\
\hline
ET0206	&$	-1.33	$&	0.17	&	24	&	41	&	16	&	0.07	&$	-0.13	$&	0.09	&	1	&	0.38	&$	-0.04	$&	0.41	\\
ET0232	&$	-1.00	$&	0.18	&	24	&	41	&	16	&	0.07	&$	-0.26	$&	0.10	&	1	&	0.30	&$	-0.80	$&	0.32	\\
ET0236	&$	-2.41	$&	0.21	&	16	&	29	&	8	&	0.08	&$	0.28	$&	0.22	&	1	&	0.74	&$	<0.72	$&	...	\\
ET0237	&$	-1.61	$&	0.18	&	20	&	36	&	16	&	0.09	&$	0.05	$&	0.11	&	1	&	0.36	&$	-0.46	$&	0.38	\\
ET0238	&$	-1.57	$&	0.17	&	24	&	38	&	16	&	0.08	&$	-0.09	$&	0.10	&	1	&	0.30	&$	-0.48	$&	0.34	\\
ET0239	&$	-2.26	$&	0.21	&	23	&	41	&	13	&	0.09	&$	0.30	$&	0.15	&	1	&	0.22	&$	0.50	$&	0.26	\\
ET0240	&$	-1.15	$&	0.17	&	24	&	37	&	15	&	0.08	&$	-0.31	$&	0.14	&	1	&	0.20	&$	-0.56	$&	0.23	\\
ET0241	&$	-1.41	$&	0.17	&	26	&	43	&	14	&	0.10	&$	-0.12	$&	0.11	&	1	&	0.30	&$	-0.24	$&	0.34	\\
ET0242	&$	-1.32	$&	0.17	&	22	&	34	&	19	&	0.07	&$	-0.22	$&	0.09	&	1	&	0.30	&$	0.00	$&	0.33	\\
ET0244	&$	-1.24	$&	0.17	&	24	&	34	&	17	&	0.11	&$	-0.13	$&	0.12	&	1	&	0.52	&$	0.18	$&	0.54	\\
ET0270	&$	-1.56	$&	0.16	&	23	&	35	&	17	&	0.08	&$	-0.08	$&	0.13	&	1	&	0.52	&$	-0.28	$&	0.54	\\
ET0275	&$	-1.21	$&	0.16	&	18	&	30	&	5	&	0.21	&$	-0.33	$&	0.23	&	1	&	0.42	&$	-0.60	$&	0.44	\\
ET0299	&$	-1.83	$&	0.18	&	18	&	29	&	7	&	0.13	&$	0.11	$&	0.17	&	1	&	9.21	&$	<-0.38	$&	...	\\
ET0300	&$	-1.39	$&	0.20	&	11	&	22	&	8	&	0.09	&$	-0.01	$&	0.13	&	1	&	8.59	&$	<-0.56	$&	...	\\
ET0317	&$	-1.69	$&	0.19	&	25	&	45	&	15	&	0.07	&$	0.02	$&	0.11	&	1	&	0.26	&$	0.00	$&	0.30	\\
ET0320	&$	-1.71	$&	0.21	&	27	&	47	&	13	&	0.10	&$	0.02	$&	0.12	&	1	&	0.20	&$	-0.14	$&	0.24	\\
ET0321	&$	-1.93	$&	0.18	&	25	&	39	&	12	&	0.12	&$	0.01	$&	0.14	&	1	&	0.28	&$	-0.20	$&	0.31	\\
ET0322	&$	-2.04	$&	0.27	&	25	&	40	&	10	&	0.13	&$	0.15	$&	0.17	&	1	&	0.44	&$	0.30	$&	0.46	\\
ET0327	&$	-1.32	$&	0.16	&	26	&	38	&	13	&	0.08	&$	-0.19	$&	0.10	&	1	&	0.32	&$	-0.46	$&	0.35	\\
ET0330	&$	-2.00	$&	0.24	&	25	&	37	&	10	&	0.09	&$	0.14	$&	0.13	&	1	&	0.18	&$	-0.34	$&	0.23	\\
ET0339	&$	-1.08	$&	0.14	&	27	&	34	&	18	&	0.09	&$	-0.30	$&	0.12	&	1	&	0.18	&$	-0.24	$&	0.21	\\
ET0342${^{(a)}}$	&$	-1.35	$&	0.20	&	10	&	17	&	...	&	...	&$	...	$&	...	&	...	&	...	&$	...	$&	...	\\
ET0350	&$	-1.90	$&	0.21	&	18	&	30	&	7	&	0.14	&$	0.18	$&	0.22	&	1	&	0.36	&$	-0.44	$&	0.40	\\
ET0354	&$	-1.07	$&	0.20	&	20	&	35	&	13	&	0.11	&$	-0.17	$&	0.13	&	1	&	0.40	&$	-0.18	$&	0.43	\\
ET0363	&$	-1.28	$&	0.17	&	16	&	30	&	11	&	0.12	&$	-0.09	$&	0.15	&	1	&	0.34	&$	-0.46	$&	0.37	\\
ET0369	&$	-2.35	$&	0.20	&	27	&	42	&	12	&	0.08	&$	0.21	$&	0.15	&	1	&	0.18	&$	0.06	$&	0.22	\\
ET0373	&$	-1.96	$&	0.21	&	19	&	35	&	10	&	0.11	&$	0.22	$&	0.19	&	1	&	0.46	&$	0.32	$&	0.49	\\
ET0376	&$	-1.17	$&	0.17	&	23	&	33	&	15	&	0.10	&$	-0.20	$&	0.13	&	1	&	0.26	&$	-0.66	$&	0.29	\\
ET0378	&$	-1.18	$&	0.15	&	23	&	34	&	18	&	0.08	&$	-0.31	$&	0.12	&	1	&	0.22	&$	-0.28	$&	0.26	\\
ET0379	&$	-1.65	$&	0.18	&	22	&	37	&	13	&	0.06	&$	-0.16	$&	0.09	&	1	&	0.26	&$	-0.20	$&	0.29	\\
ET0382	&$	-1.74	$&	0.23	&	23	&	38	&	10	&	0.08	&$	-0.06	$&	0.11	&	1	&	0.48	&$	0.60	$&	0.51	\\
ET0384	&$	-1.46	$&	0.22	&	18	&	30	&	11	&	0.10	&$	-0.04	$&	0.12	&	1	&	0.36	&$	-0.28	$&	0.39	\\
ET0389	&$	-1.60	$&	0.22	&	21	&	33	&	15	&	0.09	&$	-0.04	$&	0.11	&	1	&	0.26	&$	-0.56	$&	0.30	\\
ET0392	&$	-1.48	$&	0.20	&	15	&	26	&	12	&	0.11	&$	-0.01	$&	0.14	&	1	&	0.46	&$	-0.02	$&	0.48	\\
\hline																									
ET0034	&$	-2.41	$&	0.09	&	29	&	47	&	8	&	0.05	&$	0.32	$&	0.07	&	1	&	0.42	&$	0.10	$&	0.43	\\
ET0135	&$	-2.32	$&	0.06	&	29	&	46	&	12	&	0.04	&$	0.10	$&	0.06	&	1	&	0.20	&$	0.18	$&	0.21	\\
ET0147	&$	-1.39	$&	0.08	&	16	&	24	&	16	&	0.10	&$	-0.31	$&	0.14	&	0	&	...	&$	<-0.80	$&	...	\\
ET0156	&$	-1.37	$&	0.06	&	23	&	35	&	23	&	0.06	&$	-0.08	$&	0.11	&	1	&	0.50	&$	-0.42	$&	0.50	\\
ET0223	&$	-1.91	$&	0.07	&	16	&	30	&	11	&	0.10	&$	0.25	$&	0.11	&	1	&	0.56	&$	-0.22	$&	0.56	\\
ET0332	&$	-2.17	$&	0.08	&	22	&	36	&	10	&	0.09	&$	0.22	$&	0.15	&	1	&	0.24	&$	0.10	$&	0.25	\\
ET0351	&$	-1.27	$&	0.09	&	17	&	33	&	16	&	0.07	&$	-0.03	$&	0.11	&	1	&	0.32	&$	0.32	$&	0.33	\\
ET0359	&$	-2.67	$&	0.11	&	18	&	31	&	6	&	0.17	&$	0.42	$&	0.19	&	1	&	0.46	&$	0.46	$&	0.47	\\
ET0375	&$	-2.28	$&	0.10	&	26	&	41	&	7	&	0.14	&$	0.26	$&	0.16	&	1	&	0.46	&$	-0.14	$&	0.47	\\
ET0388	&$	-1.97	$&	0.08	&	14	&	25	&	10	&	0.13	&$	0.23	$&	0.15	&	0	&	...	&$	<-0.34	$&	...	\\
ET0393	&$	-2.12	$&	0.09	&	18	&	30	&	10	&	0.13	&$	0.19	$&	0.15	&	0	&	...	&$	<0.02	$&	...	\\
ET0394	&$	-1.39	$&	0.08	&	19	&	33	&	18	&	0.06	&$	-0.23	$&	0.10	&	1	&	0.36	&$	-0.08	$&	0.37	\\
ET0396	&$	-1.59	$&	0.06	&	17	&	28	&	14	&	0.07	&$	-0.04	$&	0.08	&	0	&	...	&$	<0.98	$&	...	\\
ET0398	&$	-1.29	$&	0.08	&	19	&	29	&	22	&	0.09	&$	-0.11	$&	0.12	&	1	&	0.40	&$	-0.50	$&	0.41	\\
ET0402	&$	-1.30	$&	0.08	&	18	&	33	&	18	&	0.11	&$	-0.10	$&	0.13	&	1	&	0.48	&$	0.16	$&	0.49	\\
ET0408	&$	-1.49	$&	0.07	&	14	&	29	&	12	&	0.11	&$	-0.15	$&	0.12	&	0	&	...	&$	<-0.22	$&	...	\\

\hline
\multicolumn{7}{l}{$^{(a)}$ S/N too low.}\\
\end{tabular}
\end{sidewaystable*}

\clearpage
\pagebreak

\begin{table}
\caption{Velocity measurements for the sample of stars, over a period ranging from 2003-2013.}     
\label{tab4:vr}   
\centering
\footnotesize
\tabcolsep=0.12cm
\begin{tabular}{ l r r r r l }
\hline\hline       
Star	&	$v_{r,1}$ (1)	&	$v_{r,2}$ (2)	&	$v_{r,3}$ (3)	&	$v_{r,4}$ (4)	&	Comment	\\
	&	(km/s)	&	(km/s)	&	(km/s)	&	(km/s)	&		\\
\hline
ET0024	&	112.9	&	112.7	&	113.3	&	110.8	&		\\
ET0026	&	98.4	&	99.2	&	97.1	&	97.4	&		\\
ET0027	&	112.3	&	114.3	&	111.8	&	113.0	&		\\
ET0028	&	120.2	&	118.7	&	119.8	&	121.5	&		\\
ET0031	&	115.1	&	115.4	&	113.3	&	116.6	&		\\
ET0033	&	108.7	&	107.5	&	106.9	&	108.8	&		\\
ET0043	&	110.2	&	109.9	&	110.2	&	110.3	&		\\
ET0048	&	120.5	&	120.5	&	121.4	&	121.7	&		\\
ET0051	&	108.5	&	110.3	&	107.0	&	108.7	&		\\
ET0054	&	103.5	&	102.9	&	102.1	&	102.9	&		\\
ET0057	&	99.3	&	98.0	&	97.3	&	100.6	&		\\
ET0059	&	115.9	&	115.7	&	118.2	&	118.7	&		\\
ET0060	&	96.6	&	96.3	&	97.1	&	97.2	&		\\
ET0062	&	105.0	&	105.1	&	103.5	&	105.2	&		\\
ET0063	&	109.4	&	108.8	&	108.5	&	109.9	&		\\
ET0064	&	113.3	&	114.2	&	113.4	&	114.8	&		\\
ET0066	&	118.6	&	116.9	&	118.2	&	121.1	&		\\
ET0067	&	99.8	&	99.8	&	100.4	&	100.6	&		\\
ET0069	&	101.9	&	101.4	&	100.2	&	102.4	&		\\
ET0071	&	106.5	&	104.7	&	106.8	&	106.3	&		\\
ET0073	&	121.9	&	122.3	&	123.1	&	124.3	&		\\
ET0083	&	122.2	&	122.9	&	121.6	&	123.2	&		\\
ET0094	&	119.4	&	106.8	&	116.6	&	110.3	&	Likely binary	\\
ET0095	&	109.5	&	111.0	&	108.4	&	112.1	&		\\
ET0097$^{(a)}$	&	105.4	&	112.6	&	106.8	&	111.4	&	Possible binary	\\
ET0103	&	116.0	&	115.0	&	116.8	&	118.0	&		\\
ET0104	&	110.6	&	...	&	110.2	&	112.5	&		\\
ET0109	&	102.5	&	101.4	&	108.4	&	103.7	&	Possible binary	\\
ET0112	&	...	&	...	&	118.2	&	117.0	&		\\
ET0113	&	122.6	&	121.4	&	119.8	&	121.0	&		\\
ET0121	&	113.9	&	116.9	&	116.6	&	113.6	&		\\
ET0126	&	102.7	&	103.1	&	101.9	&	103.6	&		\\
ET0132	&	101.6	&	100.1	&	100.4	&	101.7	&		\\
ET0133	&	111.8	&	111.4	&	111.7	&	112.5	&		\\
ET0137	&	114.5	&	115.1	&	114.0	&	115.2	&		\\
ET0138	&	110.2	&	108.4	&	108.5	&	110.2	&		\\
ET0139	&	91.1	&	90.5	&	100.3	&	100.0	&	Likely binary	\\
ET0141	&	120.4	&	122.7	&	121.5	&	122.5	&		\\
ET0145	&	...	&	...	&	106.9	&	110.3	&		\\
ET0147	&	111.8	&	110.7	&	111.8	&	112.5	&		\\
ET0150	&	100.0	&	98.9	&	97.5	&	101.0	&		\\
ET0151	&	104.0	&	105.6	&	103.5	&	106.3	&		\\
ET0158	&	111.0	&	111.2	&	111.6	&	112.3	&		\\
\hline
\end{tabular}
\end{table}

\setcounter{table}{2}
\begin{table}
\caption{Velocity measurements, continued.} 
\label{tab4:vr2}   
\footnotesize
\centering
\tabcolsep=0.12cm
\begin{tabular}{ l r r r r l }   
\hline\hline       
Star	&	$v_{r,1}$ (1)	&	$v_{r,2}$ (2)	&	$v_{r,3}$ (3)	&	$v_{r,4}$ (4)	&	Comment	\\
	&	(km/s)	&	(km/s)	&	(km/s)	&	(km/s)	&		\\
\hline
ET0160	&	112.7	&	113.5	&	111.9	&	113.2	&		\\
ET0163	&	121.2	&	116.6	&	126.3	&	117.0	&	Likely binary	\\
ET0164	&	112.7	&	114.4	&	113.2	&	113.3	&		\\
ET0165	&	113.2	&	112.0	&	111.8	&	112.9	&		\\
ET0166	&	117.4	&	118.7	&	118.2	&	118.3	&		\\
ET0168	&	113.2	&	113.1	&	113.3	&	113.2	&		\\
ET0173	&	103.6	&	107.0	&	113.6	&	108.2	&	Likely binary	\\
ET0198	&	119.4	&	119.1	&	120.0	&	120.9	&		\\
ET0200	&	104.0	&	103.0	&	103.6	&	104.5	&		\\
ET0202	&	108.6	&	107.3	&	107.0	&	108.3	&		\\
ET0206	&	95.4	&	105.4	&	92.2	&	107.4	&	Likely binary	\\
ET0232	&	102.6	&	102.3	&	103.4	&	103.6	&		\\
ET0236	&	107.1	&	102.9	&	105.3	&	104.9	&		\\
ET0237	&	111.3	&	108.9	&	111.7	&	111.1	&		\\
ET0238	&	115.8	&	114.1	&	111.7	&	114.2	&		\\
ET0239	&	116.3	&	117.4	&	118.3	&	117.6	&		\\
ET0240	&	106.6	&	105.3	&	105.3	&	107.0	&		\\
ET0241	&	111.4	&	109.2	&	111.6	&	112.5	&		\\
ET0242	&	117.5	&	117.5	&	118.1	&	118.6	&		\\
ET0244	&	112.4	&	113.3	&	113.3	&	114.6	&		\\
ET0270	&	109.5	&	...	&	...	&	109.9	&		\\
ET0275	&	112.5	&	...	&	113.5	&	113.9	&		\\
ET0299	&	93.4	&	...	&	88.9	&	92.5	&		\\
ET0300	&	119.1	&	119.7	&	119.8	&	121.8	&		\\
ET0317	&	103.9	&	103.6	&	103.5	&	104.3	&		\\
ET0320	&	119.8	&	120.7	&	119.8	&	120.7	&		\\
ET0321	&	112.6	&	112.4	&	111.6	&	113.2	&		\\
ET0322	&	104.5	&	103.3	&	100.3	&	103.6	&		\\
ET0327	&	119.1	&	119.0	&	118.4	&	119.1	&		\\
ET0330	&	114.4	&	112.4	&	113.3	&	114.0	&		\\
ET0339	&	100.6	&	98.5	&	100.2	&	101.4	&		\\
ET0342	&	119.9	&	120.3	&	118.2	&	120.0	&		\\
ET0350	&	110.9	&	110.3	&	108.5	&	110.3	&		\\
ET0354	&	105.6	&	104.6	&	105.1	&	106.3	&		\\
ET0363	&	118.4	&	119.1	&	116.5	&	118.8	&		\\
ET0369	&	103.9	&	101.2	&	105.2	&	110.2	&	Likely binary	\\
ET0373	&	130.8	&	130.5	&	129.5	&	128.0	&		\\
ET0376	&	105.8	&	106.7	&	103.6	&	105.2	&		\\
ET0378	&	109.7	&	109.0	&	107.0	&	107.3	&		\\
ET0379	&	105.5	&	105.6	&	105.2	&	105.5	&		\\
ET0382	&	103.0	&	101.9	&	102.1	&	102.2	&		\\
ET0384	&	124.6	&	124.1	&	124.9	&	126.1	&		\\
ET0389	&	115.1	&	116.7	&	114.9	&	115.5	&		\\
ET0392	&	124.6	&	124.9	&	124.6	&	126.4	&		\\
\hline
\multicolumn{6}{l}{{(1)} Hill et al. in prep.}\\
\multicolumn{6}{l}{(2)  \citet{Battaglia2008b}}\\
\multicolumn{6}{l}{(3) \citet{Skuladottir2015b}}\\
\multicolumn{6}{l}{(4) This work}\\
\multicolumn{6}{l}{(a) $v_{r,5}=109.0$~km/s, from \citet{Skuladottir2015a}}\\
\end{tabular}
\end{table}

\clearpage
\pagebreak

\begin{table}
\caption{Linelist. Horizontal lines separate which lines are fitted together. Lines that contribute $\lesssim30\%$ to the depth are in italics.} 
\label{tab4:linelisti}   
\small
\centering
\begin{tabular}{| l c c r | l c c r  | l c c r | }
\hline\hline      
El. & $\lambda$	&	$\chi_{ex}$	&	$\log (gf)$  & El. & $\lambda$	&	$\chi_{ex}$	&	$\log (gf)$    & El. & $\lambda$	&	$\chi_{ex}$	&	$\log (gf)$\\
\hline
\hline
&\multicolumn{2}{c}{Ti lines}&							&	&\multicolumn{2}{c}{Fe lines}&							&	Fe I	&	4907.732	&	3.430	&$	-1.840	$\\	\cline{1-4}	\cline{5-8}	
Ti II	&	4708.662	&	1.237	&$	-2.340	$&	Fe I	&	4741.529	&	2.831	&$	-1.765	$&	\textit{Fe I}	&	4907.967	&	4.186	&$	-1.990	$\\		\cline{5-8}	
\textit{Ti  I}	&	4708.951	&	2.160	&$	-1.053	$&	Fe I	&	4745.800	&	3.654	&$	-1.270	$&	\textit{Fe I}	&	4908.031	&	4.218	&$	-2.077	$\\	\cline{1-4}	\cline{5-8}	\cline{9-12}
Ti I	&	4715.302	&	0.048	&$	-2.680	$&	Fe I	&	4757.578	&	3.274	&$	-2.040	$&	Fe I	&	4909.383	&	3.929	&$	-1.231	$\\	\cline{1-4}	\cline{5-8}	\cline{9-12}
Ti I	&	4722.606	&	1.053	&$	-1.330	$&	Fe I	&	4765.457	&	4.076	&$	-1.938	$&	Fe I	&	4910.017	&	3.397	&$	-1.408	$\\			\cline{9-12}
Ti I	&	4723.163	&	1.067	&$	-1.335	$&	\textit{Fe I}	&	4765.480	&	1.608	&$	-4.010	$&	Fe I	&	4910.325	&	4.191	&$	-0.459	$\\	\cline{1-4}	\cline{5-8}	
Ti I	&	4731.165	&	2.175	&$	-0.134	$&	Fe I	&	4768.320	&	3.686	&$	-1.070	$&	Fe I	&	4910.565	&	4.218	&$	-0.433	$\\	\cline{1-4}		\cline{9-12}
\textit{Ti I}	&	4742.106	&	2.154	&$	-0.670	$&	Fe I	&	4768.396	&	2.940	&$	-2.261	$&	Fe I	&	4917.230	&	4.191	&$	-1.180	$\\		\cline{5-8}	\cline{9-12}
Ti I	&	4742.303	&	1.460	&$	-1.120	$&	Fe I	&	4771.697	&	2.198	&$	-3.234	$&	\textit{Fe I}	&	4917.876	&	3.017	&$	-3.876	$\\	\cline{1-4}	\cline{5-8}	
Ti I	&	4742.789	&	2.236	&$	0.210	$&	Fe I	&	4779.439	&	3.415	&$	-2.020	$&	Fe I	&	4918.012	&	4.230	&$	-1.360	$\\	\cline{1-4}	\cline{5-8}	\cline{9-12}
Ti I	&	4758.118	&	2.249	&$	0.425	$&	\textit{Fe I}	&	4787.495	&	3.018	&$	-4.162	$&	Fe I	&	4918.994	&	2.865	&$	-0.342	$\\	\cline{1-4}		\cline{9-12}
\textit{Ti I}	&	4758.901	&	0.836	&$	-2.170	$&	Fe I	&	4787.827	&	2.998	&$	-2.530	$&	Fe I	&	4920.502	&	2.832	&$	0.068	$\\		\cline{5-8}	\cline{9-12}
Ti I	&	4759.270	&	2.256	&$	0.514	$&	Fe I	&	4788.757	&	3.237	&$	-1.763	$&	Fe I	&	4924.770	&	2.279	&$	-2.241	$\\	\cline{1-4}	\cline{5-8}	\cline{9-12}
Ti II	&	4763.881	&	1.221	&$	-2.360	$&	Fe I	&	4789.651	&	3.546	&$	-0.958	$&	\textit{Fe I}	&	4930.053	&	3.301	&$	-3.255	$\\	\cline{1-4}	\cline{5-8}	
Ti II	&	4764.524	&	1.237	&$	-2.950	$&	Fe I	&	4800.649	&	4.143	&$	-1.029	$&	Fe I	&	4930.315	&	3.960	&$	-1.201	$\\	\cline{1-4}	\cline{5-8}	\cline{9-12}
Ti I	&	4771.099	&	0.826	&$	-2.380	$&	\textit{Fe I}	&	4802.875	&	3.695	&$	-2.027	$&	\textit{Fe I}	&	4932.978	&	3.274	&$	-3.222	$\\	\cline{1-4}		
Ti II	&	4775.635	&	1.243	&$	-2.810	$&	Fe I	&	4802.880	&	3.642	&$	-1.514	$&	\textit{Fe I}	&	4933.191	&	4.191	&$	-1.659	$\\	\cline{1-4}	\cline{5-8}	
Ti I	&	4778.255	&	2.236	&$	-0.220	$&	Fe I	&	4809.933	&	4.178	&$	-1.891	$&	Fe I	&	4933.293	&	3.301	&$	-2.188	$\\	\cline{1-4}		
Ti II	&	4779.985	&	2.048	&$	-1.260	$&	Fe I	&	4809.938	&	3.573	&$	-2.720	$&	Fe I	&	4933.341	&	4.231	&$	-0.817	$\\	\cline{1-4}	\cline{5-8}	\cline{9-12}
Ti I	&	4781.711	&	0.848	&$	-1.960	$&	Fe I	&	4817.778	&	2.223	&$	-3.530	$&	Fe I	&	4938.174	&	3.943	&$	-0.906	$\\	\cline{1-4}		\cline{9-12}
Ti I	&	4796.207	&	2.333	&$	-0.665	$&	\textit{Fe I}	&	4817.843	&	4.154	&$	-2.636	$&	Fe II	&	4938.817	&	8.035	&$	-4.721	$\\		\cline{5-8}	\cline{9-12}
\textit{Ti I}	&	4796.353	&	1.879	&$	-2.656	$&	Fe I	&	4834.507	&	2.424	&$	-3.410	$&	\textit{Fe I}	&	4939.239	&	4.608	&$	-4.088	$\\	\cline{1-4}	\cline{5-8}	
\textit{Ti I}	&	4797.975	&	2.334	&$	-0.768	$&	Fe I	&	4839.544	&	3.267	&$	-1.822	$&	Fe I	&	4939.241	&	4.154	&$	-0.829	$\\			\cline{9-12}
Ti II	&	4798.521	&	1.080	&$	-2.680	$&	\textit{Fe I}	&	4839.886	&	4.733	&$	-1.329	$&	Fe I	&	4939.687	&	0.859	&$	-3.340	$\\	\cline{1-4}	\cline{5-8}	\cline{9-12}
Ti I	&	4801.901	&	0.818	&$	-3.111	$&	Fe I	&	4841.663	&	3.301	&$	-2.893	$&	Fe I	&	4946.387	&	3.368	&$	-1.170	$\\			\cline{9-12}
Ti I	&	4801.948	&	0.826	&$	-3.254	$&	Fe I	&	4841.785	&	4.191	&$	-1.880	$&	Fe I	&	4950.106	&	3.417	&$	-1.670	$\\	\cline{1-4}	\cline{5-8}	\cline{9-12}
Ti II	&	4805.085	&	2.061	&$	-0.960	$&	\textit{Fe I}	&	4842.715	&	4.220	&$	-2.048	$&	Fe I	&	4962.572	&	4.178	&$	-1.182	$\\			\cline{9-12}
\textit{Ti I}	&	4805.415	&	2.345	&$	0.150	$&	Fe I	&	4842.788	&	4.103	&$	-1.560	$&	Fe I	&	4966.089	&	3.332	&$	-0.871	$\\	\cline{1-4}	\cline{5-8}	\cline{9-12}
Ti II	&	4806.321	&	1.084	&$	-3.380	$&	Fe I	&	4843.143	&	3.396	&$	-1.840	$&	Fe I	&	4967.897	&	4.191	&$	-0.487	$\\			\cline{9-12}
\textit{Ti I}	&	4806.759	&	0.813	&$	-3.113	$&	\textit{Fe I}	&	4843.348	&	0.000	&$	-7.894	$&	Fe I	&	4969.917	&	4.217	&$	-0.710	$\\	\cline{1-4}		\cline{9-12}
Ti I	&	4820.411	&	1.502	&$	-0.441	$&	\textit{Fe I}	&	4843.386	&	3.573	&$	-2.859	$&	&\multicolumn{2}{c}{Zn lines}&							\\	\cline{1-4}	\cline{5-8}	\cline{9-12}
Ti I	&	4840.874	&	0.900	&$	-0.509	$&	Fe I	&	4859.741	&	2.875	&$	-0.764	$&	Zn I	&	4722.153	&	4.030	&$	-0.338	$\\	\cline{1-4}	\cline{5-8}	\cline{9-12}
Ti II	&	4849.169	&	1.131	&$	-3.000	$&	\textit{Fe I}	&	4862.538	&	4.154	&$	-2.419	$&	Zn I	&	4810.528	&	4.078	&$	-0.137	$\\	\cline{1-4}		\cline{9-12}
\textit{Ti II}	&	4855.905	&	3.095	&$	-1.470	$&	Fe I	&	4862.599	&	4.154	&$	-1.498	$&	\multicolumn{4}{c}{} 							\\		\cline{5-8}	
Ti I	&	4856.010	&	2.256	&$	0.440	$&	Fe I	&	4863.644	&	3.430	&$	-1.663	$&	\multicolumn{4}{c}{} 							\\	\cline{1-4}		
Ti II	&	4865.611	&	1.116	&$	-2.790	$&	\textit{Fe I}	&	4863.777	&	3.039	&$	-3.252	$&	\multicolumn{4}{c}{} 							\\		\cline{5-8}	
\textit{Ti I}	&	4865.781	&	2.578	&$	-0.398	$&	Fe I	&	4867.529	&	1.608	&$	-4.700	$&	\multicolumn{4}{c}{} 							\\	\cline{1-4}		
Ti I	&	4870.126	&	2.249	&$	0.518	$&	\textit{Fe I}	&	4867.640	&	3.267	&$	-4.007	$&	\multicolumn{4}{c}{} 							\\	\cline{1-4}	\cline{5-8}	
Ti I	&	4885.079	&	1.887	&$	0.358	$&	Fe I	&	4871.318	&	2.865	&$	-0.363	$&	\multicolumn{4}{c}{} 							\\	\cline{1-4}	\cline{5-8}	
Ti I	&	4909.098	&	0.826	&$	-2.401	$&	Fe I	&	4872.138	&	2.882	&$	-0.567	$&	\multicolumn{4}{c}{} 							\\	\cline{1-4}	\cline{5-8}	
Ti II	&	4911.193	&	3.124	&$	-0.610	$&	Fe I	&	4877.604	&	2.998	&$	-3.150	$&	\multicolumn{4}{c}{} 							\\	\cline{1-4}		
Ti I	&	4913.614	&	1.873	&$	0.160	$&	\textit{Fe I}	&	4877.789	&	3.274	&$	-4.129	$&	\multicolumn{4}{c}{} 							\\	\cline{1-4}	\cline{5-8}	
Ti I	&	4919.860	&	2.160	&$	-0.225	$&	Fe I	&	4882.143	&	3.417	&$	-1.640	$&	\multicolumn{4}{c}{} 							\\	\cline{1-4}	\cline{5-8}	
Ti I	&	4926.148	&	0.818	&$	-2.170	$&	Fe I	&	4885.430	&	3.882	&$	-0.971	$&	\multicolumn{4}{c}{} 							\\	\cline{1-4}	\cline{5-8}	
Ti I	&	4928.336	&	2.154	&$	0.050	$&	Fe I	&	4886.332	&	4.154	&$	-0.613	$&	\multicolumn{4}{c}{} 							\\		\cline{5-8}	
\textit{Ti I}	&	4928.339	&	2.267	&$	-0.929	$&	Fe I	&	4889.001	&	2.198	&$	-2.462	$&	\multicolumn{4}{c}{} 							\\	\cline{1-4}		
Ti I	&	4937.726	&	0.813	&$	-2.254	$&	\textit{Fe I}	&	4889.102	&	3.884	&$	-1.170	$&	\multicolumn{4}{c}{} 							\\	\cline{1-4}	\cline{5-8}	
\textit{Ti I}	&	4940.969	&	1.981	&$	-1.880	$&	Fe I	&	4892.859	&	4.217	&$	-1.290	$&	\multicolumn{4}{c}{} 							\\		\cline{5-8}	
Ti I	&	4941.303	&	0.826	&$	-2.929	$&	Fe I	&	4903.310	&	2.882	&$	-0.926	$&	\multicolumn{4}{c}{} 							\\		\cline{5-8}	
Ti I	&	4941.571	&	2.160	&$	-1.010	$&	Fe I	&	4905.133	&	3.928	&$	-2.050	$&	\multicolumn{4}{c}{} 							\\	\cline{1-4}		
Ti I	&	4947.973	&	0.818	&$	-2.882	$&	\textit{Fe I}	&	4905.219	&	3.301	&$	-3.381	$&	\multicolumn{4}{c}{} 							\\			
\textit{Ti I}	&	4948.162	&	3.441	&$	-3.551	$&	\textit{Fe II}	&	4905.339	&	0.301	&$	-8.072	$&	\multicolumn{4}{c}{} 							\\	\cline{1-4}	\cline{5-8}	
Ti I	&	4967.296	&	0.000	&$	-3.744	$&			\multicolumn{4}{c}{} 					&	\multicolumn{4}{c}{} 							\\	\cline{1-4}		

\end{tabular}
\end{table}

\end{document}